\documentclass[12pt]{article}
\newcommand{\qed}{\nolinebreak \hfill{$\Box $} \par\vspace{0.5\parskip}\setcounter{equation}{0}}

\usepackage{txfonts}

\usepackage{url}
\usepackage{txfonts}
\usepackage{graphicx}
\usepackage{pdfpages}

\begin{document}

\begin{center}

{\large \bf Solving Differential Equations by Differentiating}
\vskip .5cm

{\it Alberto Contreras-Cristan, Jose Gonzalez-Barrios  and Raul Rueda}

Instituto de Investigaciones en Matem\'aticas Aplicadas y en Sistemas.

Universidad Nacional Aut\'onoma de M\'exico
\end{center}

\begin{center}
    {\large\bf Abstract}
\end{center}

{\it In this work,  we illustrate and explore the use of Taylor series as solutions of differential equations.
For a large a number of classes of differential equations in the literature,   there are plenty of sources 
where the well known {\it Taylor Series  Method}  is used  to approximate the solution,   but  here we are  
focused in seeing the Taylor series as a solution,   which in turn prompt us to find the recursions defining 
the coefficients in the series.   Because these recursions are found by differentiating,  instead of integrating the differential equation,
it is not difficult to prove that the resulting series is a solution.   In the case where the series does not have a 
closed analytic form or it is not a known function,  Cauchy-Hadamard  theorems can be used to find  the radius of convergence and then
the series is a solution for the differential equation, in the domain where it converges.} \newline \newline
{KEYWORDS:} Univariate Differential Equations; Partial Differential
Equations; Taylor's Theorem; Radius of Convergence of Taylor's and  Maclaurin's Series.

\vskip0.5cm
\noindent {\bf 1.-  Introduction} Differential equations and their solution have been  important
research topics   in ma\-thematical applications. The mathematical solution of many problems in areas
such as Physics, Chemistry, Engineering,  among others, depends on solving adequate
differential equations, that is  why, for several centuries, many mathematicians have put special attention in defining and solving them. Solving differential equations has been linked with the creation of areas such as differential calculus, integral  calculus, and many others. However, the solution of differential equations has been always connected with the problem of integrating functions, which in some  cases is not easy to solve, for example,  it is well known that the solutions of some integrals have no closed analytic forms, that is  the case  of the integral of the Gaussian density function. This issue had already been noted by Euler in 1758 (see \cite{FA}) and it motivates the idea of iteratively differentiating  the differential equation and  using its initial values,  in order to compute a Taylor series expansion for its solution.  The well-known Taylor series method (\cite{B}, \cite{CH} and \cite{GR}),  has been used to numerically approximate the solution to ordinary differential equations, boundary
value problems and partial differential equations that appear in a wide range of applications.  In the context of integral equations (\cite{YA}), integro-differential equations (\cite{ME}), differential-difference equations (\cite{GU}) and fractional integro-differential equations (\cite{HU}),  a Taylor matrix method is used to propose a finite Taylor polynomial as an approximate solution, which, for some examples in  \cite{YA}  and   \cite{ME},  yields an exact (analytic) solution.   The present work aims to explore
the use of Taylor series as solutions of some differential equations,  whenever the coefficients of the series can be identified by solving a recursive relation.  A similar scheme is followed in \cite{CHa} for the case of multivariate functions.  If the series converges to an unknown function, the Cauchy-Hadamard theorems can be used to find the radius of convergence, and then the series is a solution for the differential equation,  in the domain where it converges.

In this paper,  if $f:\rightarrow \mathbb{R}$ is a $k$-differentiable function, where $k$ is a nonnegative integer, we denote the $k^{\mbox{th}}$ derivative of $f$ by
$$f^{(k)}(x)=d^{k}f(x)/dx^k.$$

Now we recall {\bf Taylor's theorem.}

\noindent {\bf Theorem}\label{taylorT}
{\it Let $k\geq 1$ be an integer, let $x_0\in\mathbb{R}$ and let $f:\mathbb{R}\rightarrow\mathbb{R}$ be a $k$ times differentiable function at $x_0$. Then there exists  a function $h_k:\mathbb{R}\rightarrow\mathbb{R}$ such that

\begin{equation}\label{taylor}
f(x) = \sum_{i=0}^{k} \frac{f^{(i)}(x_0)}{i!} (x-x_0)^i + h_k(x)(x-x_0)^k \quad\mbox{for every}\quad x\in \mathbb{R},
\end{equation}
where $\lim_{x\rightarrow x_0}h_k(x)=0$, and this is usually called the {\bf Peano form of the reminder}.}

If we let $P_{k}(x) = \sum_{i=0}^{k} \frac{f^{(i)}(x_0)}{i!} (x-x_0)^i$ then $P_k$ is the
{\bf $k$-th order Taylor polynomial}. So, from equation (\ref{taylor}), $f(x) = P_k(x) + h_k(x)(x-x_0)^k$ for every $x\in\mathbb{R}$. Note that $f(x)=P_{k}(x)$ for every $x\in\mathbb{R}$ if and only if $f$ is a polynomial of degree $j\leq k$.

\vskip0.1cm
\noindent {\bf Definition}\label{analytic}
{\it Let $I\subset \mathbb{R}$ be an open interval and let $f:I\rightarrow \mathbb{R}$ be a function. Then
$f$ is {\bf a real analytic} if it is locally defined by a convergent power series.
This means for every $x_0\in I$, there exists $r>0$ and a sequence of coefficients $\{c_k\}_{k\geq 0}\subset \mathbb{R}$, such that $(x_0-r,x_0+r)\subset I$ and
\begin{equation}\label{series}
f(x)=\sum_{k=0}^{\infty} c_k(x-x_0)^k \quad \mbox{for every} \quad x\in (x_0-r,x_0+r).
\end{equation}}
In general, the radius of convergence $r$ of a power series can be computed from Cauchy-Hadamard's formula given by
\begin{equation}\label{radius}
\frac{1}{r} =\limsup_{k\rightarrow\infty} |c_k|^{\frac{1}{k}}.
\end{equation}
For Taylor polynomials we know that $c_k=\frac{f^{(k)}(x_0)}{k!}$ for every integer $k\geq 0$.

\vskip 1cm
\noindent{\bf 2.- Univariate Examples}
\vskip0.1cm
\noindent {\bf 2.1  The Gaussian Distribution Function.}
\vskip0.1cm

\noindent Let $\Phi^{(1)}(x)=D e^{-\frac{x^2}{2}}$ for $x\in\mathbb{R}$, with $D=\frac{1}{\sqrt{2\pi}}$, and with initial condition $\Phi(0)=\frac{1}{2}$. Then we have that $\Phi^{(1)}(0)=D$. 
\vskip0.1cm
\noindent {\bf Lemma 1.}
{\it Let $\Phi$ be the solution of the above differential equation, then the recursion to find the Taylor series  for $\Phi,$ around $x=0,$ 
is defined for $k\geq2$ as
\begin{equation}\label{itereqn}
\Phi^{(k)}(x)=\Phi^{(1)}(x)\cdot P_k(x), \quad\mbox{where}\quad P_k(x)=-x\cdot P_{k-1}(x)+P_{k-1}^{(1)}(x),
\end{equation}
with  $P_1(x) = 1$ and  $P_2(x)=-x$.}
\vskip0.1cm

\noindent {\bf Proof:}
Differentiating   $\Phi^{(1)}(x)$  we get
$$ \Phi^{(2)}(x)=-D x e^{-\frac{x^2}{2}}=D e^{-\frac{x^2}{2}}[-x]=\Phi^{(1)}(x)[-x]=\Phi^{(1)}(x)\cdot P_2(x),$$
where $P_2(x)=-x$ is a polynomial. So, $ \Phi^{(2)}(0)=\Phi^{(1)}(0)\cdot P_2(0)=0$. For $k=3$ we obtain
$$ \Phi^{(3)}(x)= - D e^{-\frac{x^2}{2}} +D x^2  e^{-\frac{x^2}{2}}=\Phi^{(1)}(x)[x^2-1]=\Phi^{(1)}(x)\cdot P_3(x),$$
where $P_3(x)=x^2-1= -x\cdot P_2(x)+P_2^{(1)}(x)$ using the chain rule. Then $ \Phi^{(3)}(0)=\Phi^{(1)}(0)\cdot P_3(0)=-D$.
For $k=4$ we get
$$\Phi^{(4)}(x)=D 2x e^{-\frac{x^2}{2}}+ D x e^{-\frac{x^2}{2}}-D x^3 e^{-\frac{x^2}{2}}
=\Phi^{(1)}(x)[-x^3+3x]=\Phi^{(1)}(x)\cdot P_4(x),$$
where $P_4(x)=-x^3+3x= -x\cdot P_3(x)+P_3^{(1)}(x)$ using the chain rule. So, $\Phi^{(4)}(0)=\Phi^{(1)}(0)\cdot P_4(0)=0$.
Therefore, using the chain rule, we have an iterative expression in terms of $k$ for $\Phi^{(k)}(x)$ given by (\ref{itereqn})
hence we have finished the proof. \qed

For example, since $P_4(x)=-x^3+3x$ then by (\ref{itereqn}), 
$P_5(x)=-x\cdot P_{4}(x)+P_{4}^{(1)}(x)=-x\cdot(-x^3+3x)+(-3x^2+3)=x^4-x^3-6x^2+3$, so, $\Phi^{(5)}(0)\cdot P_5(0)=3 D$.
Equation (\ref{itereqn}) is quite easy to program in order to obtain the values of
$\Phi^{(k)}(0)$, for every $0\leq k\leq n$ for every integer $n\geq 2$. Hence, we can obtain the $n^{\mbox{th}}$ order Taylor polynomial to approximate the value of $\Phi(x)$. We use the language $R$ to make a short program to approximate the values of $\Phi(x)=\Phi^{(0)}(x)$, using Taylor or Maclaurin series around $x_0=0$, which corresponds to the distribuition function of a normal or Gaussian random variable with mean  $\mu=0$ and variance $\sigma^2=1$, which is well known not to have a closed analytic form, that is, we are providing approximate values of $\Phi(x)$ by taking
$$\Phi(x)=\int_{-\infty}^{x} \frac{1}{\sqrt{2\pi}}exp^{-\frac{t^2}{2}}dt\approx 
\sum_{k=0}^{n}\frac{\Phi^{(k)}(0)}{k!} (x-0)^k$$

\begin{table}[!h] 
\caption{Approx. Values of $\Phi(x)$ for different values of $x$ and $n$ varying between $n=5$ and $n=75$}
\renewcommand{\arraystretch}{1}
\begin{tabular}{|c|c|c|c|c|c|}\hline
$\mbox{Value of}\,x$  & $n=10$ & $n=25$ & $n=50$ & $n=75$ &  $\mbox{Real Value of}\, \Phi(x)$ \\ \hline
-4    &-17.860 &-3.3786 &-0.00002701 &{\bf 0.00003167} &{\bf 0.00003167} \\ \hline
-3.6 &-6.2837 &-0.2106 &0.00015882  &{\bf 0.00015910} &{\bf 0.00015910}  \\ \hline
-2.8 &-0.4929 &0.00228 &{\bf 0.0025551} &{\bf 0.0025551}  &{\bf 0.0025551}  \\ \hline
-2.2 &-0.0268 &0.01390 &{\bf 0.0139034} &{\bf 0.0139034}  &{\bf 0.0139034} \\ \hline
-1.5 & 0.0661 &{\bf 0.0668072} &{\bf 0.0668072} &{\bf 0.0668072}  &{\bf 0.0668072} \\ \hline
-1.0 &0.15864 &{\bf 0.1586553} &{\bf 0.1586553} &{\bf 0.1586553}  &{\bf 0.1586553} \\ \hline
0    &{\bf 0.5} &{\bf 0.5}    &{\bf 0.5}       &{\bf 0.5}        &{\bf 0.5}       \\ \hline
1.0  &0.84135 &{\bf 0.8413447} &{\bf 0.8413447} &{\bf 0.8413447} &{\bf 0.8413447} \\ \hline
1.5  &0.93389 &{\bf 0.9331928}&{\bf 0.9331928}  &{\bf 0.9331928} &{\bf 0.9331928} \\ \hline
2.2  &1.02688 &0.986097   &{\bf 0.9860966 } &{\bf 0.9860966}  &{\bf 0.9860966} \\ \hline
2.8  &1.49299 &0.9977161&{\bf 0.9974449} &{\bf 0.9974449}  &{\bf 0.9974449} \\ \hline
3.6  &7.28378 &1.210616 &0.9998412  &{\bf 0.9998409 } &{\bf 0.9998409} \\ \hline
4     &18.8606 &4.378609 &1.00027    &{\bf 0.9999683 } &{\bf 0.9999683} \\ \hline
\end{tabular}
\end{table}

In  Table 1 we give the approximated values of $\Phi(x)$ for different values of $x$ varying from $x=-4$ to $x=4$ using selected values of $n$ from $n=5$ to $n=75$. The last column corresponds to the values obtained using the $R$ command ``pnorm(x,0,1)'' to obtain the real value of $\Phi(x)$. We used bold type numbers for the approximated values which coincide with the ``real'' values obtained with $R$. Note that the approximated values given for $n=5$ and $n=10$ are quite poor for $|x|>1.5$, this should not be a surprise, since the convergence of the Taylor power series is an asymptotic result. In fact for $n=75$ we obtain the convergence for $|x|\leq 4$, which is the domain of most tables in a probability or statistics book. Another important observation is that the approximated values are given almost instantly, since the computation times reported when we run our program for $n\leq 100$ are user(time)=0 , system(time)=0 and  elapsed(time)=0.

\setcounter{equation}{4}
\vskip0.1cm
\noindent {\bf 2.2  Newton's Law of Cooling.}
\vskip0.1cm

\noindent Let $T:[0,\infty)\rightarrow\mathbb{R}$ be a function which denotes
the temperature of an object, let $T(0)$ denote the initial temperature of the object. Let $T_a$ be a constant
that provides the ambient temperature. Newton provided a differential equation that
gives the value of $T(t)$ at  time $t>0$, by stating that the temperature of the object tends to reach the
ambient temperature proportionally to the difference of the temperature of the
body at time $t$ minus the ambient temperature $T_a$, following the next equation which we will write
using Lagrange's notation 
\begin{equation}\label{cooling}
T^{(1)}(t) = L \cdot (T(t)-T_a) \quad \mbox{for any} \quad t\geq 0.
\end{equation}
He proposed that the constant $L$ must be negative if $T(0)>T_a$, and $L$ must be positive if $T(0)<T_a$.

We will first use Taylor power series to solve the equation. We have to solve the equation
$T^{(1)}(t) =L\cdot T(t) -L\cdot T_a$ with initial condition $T(0)=c$, the we have that
$T^{(1)}(0)=L\cdot c- L\cdot T_a=L\cdot(c-T_a)$. Differentiating we obtain that
$$ T^{(2)}(t) =L\cdot T^{(1)}(t)= L^2\cdot T(t)-L^2 \cdot T_a.$$
Hence, $T^{(2)}(0)=L^2\cdot T(0)-L^2\cdot T_a=L^2(c-T_a)$. In general, we have that for any $k\geq 2$,
$$ T^{(k)}(t)=L\cdot T^{(k-1)}(t)= L^k(T(t)-T_a).$$
Therefore, $T^{(k)}(0)=L^k (c-T_a)$. So, using a Taylor (McLaurin) expansion around $t_0=0$
\begin{equation}\label{newton} T(t)=\sum_{k=0}^{\infty} \frac{T^{(k)}(0)}{k!} (t-0)^k 
=c +(c-Ta)\cdot \sum_{k=1}^{\infty} \frac{(L\cdot t)^k}{k!}=(c-T_a)\cdot \exp(Lt) +T_a.\end{equation}
Now we have to see that equation (\ref{newton}) is a solution of the equation (\ref{cooling}) with
inital condition $T(0)=c$. We note that from (\ref{newton}), $T(0)=c-T_a+T_a=c$. Besides,
$$ L\cdot T(t) -L\cdot T_a= L\left[ (c-T_a)\cdot\exp(Lt)+T_a\right] -L\cdot T_a=
L(c-T_a)\exp(Lt)=T^{(1)}(t).$$
So, (\ref{newton}) is a solution.

Now, let us give a ``standard'' solution of this differential equation. We have the equation
$$ \frac{dT}{dt} = L\cdot (T-T_a),$$
then $\frac{dT}{T-T_a}= L dt$ integrating we have that
$ \int \frac{dT}{T-T_a} =\int L dt$, so, $\ln|T-T_a| =L\cdot t +c$, equivalently
$$ |T-T_a|= \exp(\ln|T-T_a|)=\exp(L t+d)= M\exp(L t)\quad\mbox{for a constant}\,\, M.$$

Hence, the general solution is of the form $T(t)= M\exp(L t) + B$ where $M$ and $B$ are constants, and under
the initial condition $T(0)=c$, we get $M=(c-T_a)$ and $B=T_a$, that is, the same solution given in equation (\ref{newton}).

\vskip0.1cm
\noindent {\bf 2.3  Harmonic Oscillator.}
\vskip0.1cm
\noindent The harmonic oscillator differential equation is related to
the movement of a pendulum. Let $f(t)$ the function that indicates the angular displacement of the pendulum, let
$l$ be the length of the massless cord of the pendulum, and let $g=9.8 m/\mbox{seg}^2$ be the
acceleration due to gravity, and define $M=g/l>0$. The differential equation we want to solve is
\begin{equation}\label{pendulum}
f^{(2)}(t)=-\frac{g}{l} f(t)=-Mf(t).
\end{equation}
This equation holds for small values of $t$, but we will solve it for any $t\in\mathbb{R}$.
Let us assume that the initial conditions are $f(0)=c$ and $f^{(1)}(0)=d$, where $c,d\in\mathbb{R}$. 

\noindent {\bf Lemma 2.}
{\it  To find the solution of equation (\ref{pendulum}),  with initial conditions  $f(0) = c$ and $f^{(1)}(0) = d$ and where $c$ and $d$
are real numbers,  the recursion to find  the Taylor coefficients is given by
\begin{equation}\label{eqrec}
f^{(k)}(t)= -M f^{(k-2)}(t),  \quad \mbox{for every} \quad k \geq 4,
\end{equation}
these coefficients determine the corresponding Taylor series around $x=0$.}
\vskip0.1cm

\noindent {\bf Proof:} Using equation (\ref{pendulum}) we have that $f^{(2)}(0)= -M\cdot c$, differentiating (\ref{pendulum})
we have that
$$ f^{(3)}(t)= -M f^{(1)}(t).$$
So, $ f^{(3)}(0)=-M f^{(1)}(0)= -M\cdot d$. In general, differentiating (\ref{pendulum}) consecutively,
we have that for any $k\geq 4$
$$ f^{(k)}(t)= -M f^{(k-2)}(t).$$
Therefore, we have that $f^{(2n)}(0)= (-M)^n \cdot c$ and $f^{(2n+1)}(0)= (-M)^n \cdot d$, which holds
for any integer $n\geq 0$.  Using twice the Taylor (McLaurin) expansion  around $t_0=0,$  given in equation (\ref{series}),
we have

\begin{eqnarray}\label{solution} f(t)&=&\sum_{n=0}^{\infty} \frac{f^{(2n)}(0)}{(2n)!}t^{2n}+ 
\sum_{n=0}^{\infty} \frac{f^{(2n+1)}(0)}{(2n+1)!} t^{2n+1} \nonumber\\
&=& c\cdot \cos\left(\sqrt{M}t\right)+ \frac{d}{\sqrt{M}}\cdot \sin\left(\sqrt{M}t\right),
\end{eqnarray}
which ends the proof.  \qed

\noindent  It follows from (\ref{solution}), that $f(0)=c$. we also have that
$f^{(1)}(t)=-c\sqrt{M}\cdot \sin\left(\sqrt{M}t\right)+ d\cdot \cos\left(\sqrt{M}t\right)$.
So, $f^{(1)}(0)=d$. Besides,
$$f^{(2)}(t)=-M\left(c\cdot \cos\left(\sqrt{M}t\right)+ \frac{d}{\sqrt{M}}\cdot \sin\left(\sqrt{M}t\right)\right)=-M f(t).$$
Therefore, $f$ given in equation (\ref{solution}) is a solution of the differential equation (\ref{pendulum}),
with the given initial conditions.

\setcounter{equation}{9}
\vskip0.1cm
\noindent {\bf 2.4 A non Homogeneous Differential Equation.}
\vskip0.1cm

\noindent Let us consider the differential equation
\begin{equation}\label{nonhomog}
f^{(2)}(x)+f^{(1)}(x)=-\sin(x),
\end{equation}
with inital conditions $f(0)=c$ and $f^{(1)}(0)=d$, where $c,d\in\mathbb{R}$. 

\noindent {\bf Lemma 3.}
{\it Define the function $e_4(x)=\frac{x^4}{4!}+\frac{x^8}{8!}+\frac{x^{12}}{12!}+\cdots$, for every $x\in\mathbb{R}.$
Then,  a solution $f$ to the differential equation (\ref{nonhomog}) is given by
\begin{equation}\label{solutnonhomog} 
f(x) = c+(-d+1) e_4(x)+(d-1) e_4^{(1)}(x)+(-d)e_4^{(2)}(x)+d e_4^{(3)}(x),\,\,\,\mbox{for every}\,\,\, x\in\mathbb{R}.
\end{equation}}

\noindent {\bf Proof:}   From equation (\ref{nonhomog}) we get
that $f^{(2)}(x)=-f^{(1)}(x)-\sin(x)$, so, $f^{(2)}(0)=-f^{(1)}(0)=-d$. Differentiating equation (\ref{nonhomog})
consecutively we get
$$ f^{(3)}(x)=-f^{(2)}(x)-\cos(x), \,\,\, f^{(4)}(x)=-f^{(3)}(x)+\sin(x),$$
$$f^{(5)}(x)=-f^{(4)}(x)+\cos(x),\,\,\mbox{and}\,\, f^{(6)}(x)=-f^{(5)}(x)-\sin(x).$$
So,
$$ f^{(3)}(0)=-(-d)-1=d-1,\,\,\, f^{(4)}(0)=-(d-1)+0=-d+1,$$
$$f^{(5)}(0)=-(-d+1)+1=d,\,\,\mbox{and}\,\,
f^{(6)}(0)=-(d)-0=-d.$$
Of course, we have a periodical behavior of $f^{(j)}(0)$ given by

$$ f^{(4l+1)}(0)=d,\,\,\,f^{(4l+2)}(0)=-d,\,\,\,f^{(4l+3)}(0)=d-1,\,\,\mbox{and}\,\,f^{(4l+4)}(0)=-d+1,$$
for every $l\geq 0$. So, using a Taylor (McLaurin) series expansion we get 
\begin{eqnarray}\label{solution2}
f(x) &=& c+\sum_{l=0}^{\infty}\frac{f^{(4l+1)}(0)}{(4l+1)!}x^{4l+1} 
+\sum_{l=0}^{\infty}\frac{f^{(4l+2)}(0)}{(4l+2)!}x^{4l+2}\nonumber\\ 
&+&\sum_{l=0}^{\infty}\frac{f^{(4l+3)}(0)}{(4l+3)!}x^{4l+3}
+\sum_{l=0}^{\infty}\frac{f^{(4l+4)}(0)}{(4l+4)!}x^{4l+4}\nonumber\\
& = & c + d\left[\frac{x}{1!}+\frac{x^5}{5!}+\frac{x^9}{9!}+\cdots\right]
+(-d)\left[\frac{x^2}{2!}+\frac{x^6}{6!}+\frac{x^{10}}{10!}+\cdots\right]\nonumber\\
&+& (d-1)\left[\frac{x^3}{3!}+\frac{x^7}{7!}+\frac{x^{11}}{11!}+\cdots\right]
+(-d+1)\left[\frac{x^4}{4!}+\frac{x^8}{8!}+\frac{x^{12}}{12!}+\cdots\right]
\end{eqnarray}
By definition of $e_4(x)$ above, it follows that $0\leq e_4(x)< \exp(x)<\infty,$ for every $x\in\mathbb{R}$ and 
thus equation (\ref{solutnonhomog}) holds.  This finishes the proof. \qed

\setcounter{equation}{12}
\noindent Note also that $\sum_{j=0}^{3}e_4^{(j)}(x)=\exp(x)-1$ for every $x\in\mathbb{R}$. In fact, in general if for any positive integer $k$ we define 
\begin{equation} \label{expk}
e_k(x)=\sum_{l=1}^{\infty}\frac{x^{lk}}{(lk)!} \quad\mbox{for every}\quad x\in\mathbb{R} ,
\end{equation} 
then, from (\ref{expk}) we have that $ \left|e_k(x)\right|\leq e_k(|x|)<\exp(|x|)<\infty$ for every $x\in\mathbb{R}$, so $e_k$ is well defined,
and of course is an infinitely differentiable function, such that
$$
\sum_{j=0}^{k-1}e_k^{(j)}(x)=\exp(x)-1\quad\mbox{for every}\quad x\in\mathbb{R}.
$$
Besides, it is easy to see, that for every integer $k\geq 2$, the function $g_k(x)=e_k(x)+1$ is a different solution
from $\exp(x)$ of the differential equation $f^{(k)}(x)=f(x)$ with initial condition $f(0)=1$.

It is not difficult to see, that the function $f$ defined in equation (\ref{solution2}) is a solution of the
non homogeneous differential equation given in (\ref{nonhomog}) with initial conditions
$f(0)=c$ and $f^{(1)}(0)=d$. On the other hand, if we try to solve the differential equation using
the option {\it DSolve} given in the package ``{\it Mathematica}'', we obtain the following solution:

\begin{equation}\label{mathematicasol2}
g(x)=C_2+\frac{1}{2}\left[ -2C_1\exp(-x) +\cos(x)+\sin(x)\right],
\end{equation}
where $C_1$ and $C_2$ are constants. It is easy to check that the function $g(x)$ given
in (\ref{mathematicasol2}) is in fact a solution of the non homogeneous equation (\ref{nonhomog}).
In fact, if we define $c=g(0)=C_2-C_1+\frac{1}{2}$ and $d=g^{(1)}(0)=C_1+\frac{1}{2}$ and substitute these
values in equation (\ref{solution2}), then we obtain that $f(x)=g(x)$, because
$$
f(x) = C_2-C_1\exp(-x)+\frac{1}{2}\sin(x)+\frac{1}{2}\cos(x).
$$

\vskip0.1cm
\noindent {\bf 2.5 Homogeneous Differential Equation With Constant Coefficients.}
\vskip0.1cm

\noindent Let us consider the differential equation
\begin{equation}\label{homogconst}
f^{(2)}(x)= (c+d) f^{(1)}(x) -c\cdot d f^{(0)}(x) \quad\mbox{where} \quad c,d\in\mathbb{R},\,\, c\not=d\,\,\mbox{and}\,\,|d|<|c|,
\end{equation}
with initial conditions $f^{(0)}(0)=0$ and $f^{(1)}(0)=B,$ for some $B\in\mathbb{R}$. 

\noindent {\bf Lemma 4.}
{\it Let $f$ be a solution to the differential equation (\ref{homogconst}), then the recursion to find the Taylor coefficients 
is given by
\begin{equation}\label{soluthomogconst} 
f^{(k)}(x)=(c+d)f^{(k-1)}(x)-c\cdot d f^{(k-2)}(x), \quad\mbox{for every}\quad x\in\mathbb{R}.
\end{equation}
these coefficients determine the corresponding Taylor series around $x=0$.} 

\noindent {\bf Proof:} From equation (\ref{homogconst})
we get $f^{(2)}(0)=B(c+d)$  and  it is clear  that from equation (\ref{homogconst}),  for any $k\geq 3$ we obtain
(\ref{soluthomogconst}).   We then proceed to use (\ref{soluthomogconst}) to obtain a Taylor series expansion
for $f.$  For $k=3$ we have
$$ f^{(3)}(0)=(c+d) f^{(2)}(0)-c\cdot d f^{(1)}(0)=B(c+d)^2 -c\cdot d B=B(c^2+c\cdot d+d^2).$$
For $k=4$ we obtain
 $$f^{(4)}(0)=(c+d) f^{(3)}(0)-c\cdot d f^{(2)}(0)=B(c+d)(c^2+c\cdot d +d^2)-c\cdot d B(c+d)$$
$$= B(c^3+2c^2d+2cd^2+d3)-B(c^2d+cd^2)=B(c^3+c^2d+cd^2+d^3).$$
It is easy to see that
$$ f^{(k)}(0)=B\left(\sum_{j=0}^{k-1} c^{(k-1)-j}d^j\right).$$
Therefore,
\begin{eqnarray}\label{arreglo}
f^{(1)}(0)&=& B, \nonumber\\
f^{(2)}(0)&=& B(c+d), \nonumber\\
f^{(3)}(0)&=& B(c^2+cd+d^2), \nonumber\\
f^{(4)}(0)&=& B(c^3+c^2d+cd^2+d^3), \nonumber\\
f^{(5)}(0)&=& B (c^4+c^3d+c^2d^2+cd^3+d^4) \,\,\,\, {\rm and} \, {\rm so} \, {\rm on}
\end{eqnarray}
Now using the Taylor (McLaurin) series around $x_0=0$, we have that summing by columns in (\ref{arreglo})
\begin{eqnarray*}\label{solution3}
f(x) &=& \sum_{k=0}^{\infty} \frac{f^{(k)}(0)}{k!} x^k=0+
B\sum_{k=1}^{\infty}\frac{\left(\sum_{j=0}^{k-1} c^{(k-1)-j}d^j\right)}{k!} x^k \nonumber\\
& = & \frac{B}{c}\left(\frac{c}{c-d}\right)\left[(e^{cx}-1)-dx -\frac{(dx)^2}{2!}-\frac{(dx)^3}{3!}-\cdots\right]\nonumber\\
& = & \frac{B}{c-d}\left[e^{cx}-e^{dx}\right]
\end{eqnarray*}
It follows easily that $f$ is a solution of equation (\ref{homogconst}).    \qed

\noindent In the next  example,  we illustrate that  computation of  a McLaurin series is not allways expeditious.

\setcounter{equation}{17}
\vskip0.1cm
\noindent {\bf 2.6 A Classical Example of a Function Without a McLaurin Series.}
\vskip0.1cm

\noindent It is well known that not every function has a McLaurin or Taylor series at $x_0=0$, the best known example
is the function

\begin{equation}\label{nomclaurin}
f(x)=\left\{ \begin{array}{lcl}
\exp\left(-\frac{1}{x^2}\right) & \,\,\mbox{if}\,\, & x\not= 0 \nonumber\\
0 & \,\,\mbox{if}\,\, & x =0.
\end{array}\right.
\end{equation}
Clearly, the function in equation (\ref{nomclaurin}) is even and continuous in $\mathbb{R}\backslash \{0\}$. Besides,
$$\lim_{x\downarrow 0} \exp\left(-\frac{1}{x^2}\right)=\exp\left( -\lim_{x\downarrow 0}1/x^2\right)
=\exp(-\infty)=0.$$
So, we can extend the definition of $f$ to $\mathbb{R}$, by taking $f(0)=0$. In fact, the function $f$ around $x_0=0$
is for all purposes practically zero. For example, if we evaluate the function $f$ around $x_0=0$,
 using the language  $\mbox{\it R}$ and package $\mbox{\it Mathematica}$, 
we note that $f(1/27.29712)=\exp(-(27.29712)^2)=4.940656e-324$
and $f(1/27.29713)=\exp(-(27.29713)^2)=0$. Therefore, $f(x)\approx 0$ for every $|x|\leq \frac{1}{27.29713}$,
where the approximated values are always less than or equal $4.940656e-324$, see Figure 2 below.
Let us study carefully the values of the derivatives of $f$ at $x_0=0$, using extensions by continuity.
First, let us define $h(x)=-\frac{1}{x^2}$ for every $x\in\mathbb{R}\backslash \{0\}$. Then, $f$ is an even function
which is continuous and infinitely differentiable, its first two derivatives are given by
$$ h^{(1)}(x) = \frac{2}{x^3}, \,\, h^{(2)}(x)=-\frac{2\cdot 3}{x^4}=-\frac{3!}{x^4}\,\,\mbox{for every}\,\,
x\in\mathbb{R}\backslash \{0\}.$$
In fact, its $k^{\mbox{th}}$ derivative  is of the form
$$
h^{(k)}(x)= (-1)^{k+1} \frac{(k+1)!}{x^{k+2}}\quad\mbox{for every}\,\, k\geq 1 \,\,\mbox{and for every}\,
x\in\mathbb{R}\backslash \{0\}.
$$
Since $f(x)=\exp(h(x))=(\exp\circ h)(x)$, where $\circ$ denotes composition, 
we will obtain its first two derivatives using the chain rule. So, we have
$$ f^{(1)}(x)=h^{(1)}(x) f(x)=\frac{2}{x^3} \exp(-1/x^2),\,\,
f^{(2)}(x) =\left[-\frac{3!}{x^4}+\frac{4}{x^6}\right] f(x),$$
for every $x\in\mathbb{R}\backslash \{0\}$. The third derivative is given by
\begin{eqnarray}\label{terzader}
f^{(3)}(x)&=&f(x)\frac{d}{dx}\left[h^{(2)}(x)+\left( h^{(1)}(x)\right)^2\right]+
\left[h^{(2)}(x)+\left( h^{(1)}(x)\right)^2\right] f^{(1)}(x)\nonumber\\
&=& \exp(-1/x^2)\left[h^{(3)}(x)+3h^{(2)}(x) h^{(1)}(x)+\left( h^{(1)}(x)\right)^3\right]\nonumber\\
& = & \exp(-1/x^2)\left[ \frac{4!}{x^5} -3\frac{3!}{x^4}\frac{2}{x^3}+\frac{8}{x^9}\right].
\end{eqnarray}

\begin{figure}[h!]
\begin{center}
\includegraphics[width=6.5cm,height=6.5cm]{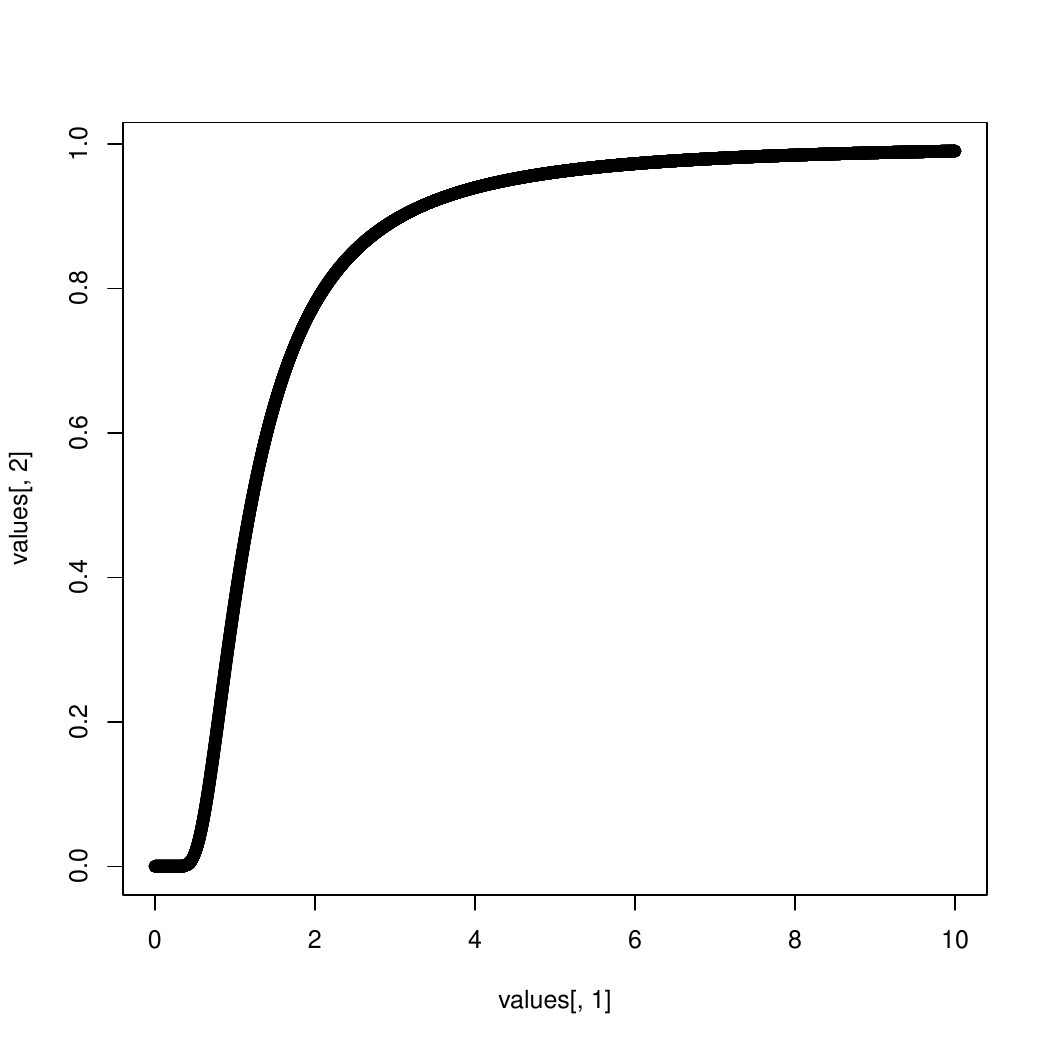}
\caption{Graph of $f(x)=\exp(-1/x^2)$ for $0<x\leq 10$}
\end{center}
\end{figure}

It is easy to see that $f^{(k)}$  can be written as $f^{(k)}(x)=f(x)\left[ g_k(x)\right]$, where $g_k(x)$ is a function that depends on the first $k$ derivatives of the function $h(x)$ defined above.
Therefore, $f^{(k+1)}(x)=f(x)\left[ g_{k}^{(1)}(x)+g_k(x)\cdot h^{(1)}(x)\right]$, which a is recursive formula for the derivatives of $f$. See Figure 1.

We can find a general expression for the $k^{\mbox{th}}$ derivative of $f$ based on the concept of partitions.
Let $n$ be a positive integer and define the set ${\cal{P}}_n$ of all integers partitions of $n$ by
$$
\hspace{-0.3cm}{\cal{P}}_n=\left\{ (n_1,\ldots ,n_k)\in \{1,\ldots ,n\}^k \,|\, n_1\geq\cdots \geq n_k\,\,\,
\mbox{for}\,\, k\geq 1, \,\, \mbox{with}\,\, \sum_{j=1}^{k} n_j=n\right\}.
$$
Using ${\cal{P}}_n$ we can write
\begin{equation}\label{kesimaderparti}
f^{(k)}(x)=f(x) \sum_{(n_1,\ldots ,n_k)\in {\cal{P}}_n} c_{(n_1,\ldots ,n_k)} 
h^{(n_1)}(x) h^{(n_2)}(x)\cdots h^{(n_k)}(x),
\end{equation}
where $c_{(n_1,\ldots ,n_k)}$ is a positive integer for every $(n_1,\ldots ,n_k)\in {\cal{P}}_n$.
Of course, the sum in equation (\ref{kesimaderparti}) coincides with $g_k(x)$ defined above.
For example, from equation (\ref{terzader}) and using the recursive equation is easy to see that
\begin{eqnarray}
f^{(4)}(x) &=& f(x) \left[ h^{(4)}(x)+4h^{(3)}(x)h^{(1)}(x)+3\left(h^{(2)}(x)\right)^2 \right. \nonumber\\
&+& 5h^{(2)}(x) \left. \left(h^{(1)}(x)\right)^2+\left(h^{(1)}(x)\right)^4\right].\nonumber
\end{eqnarray}
Here we use all the integer partitions of $n=4$ given by $(4), (3,1), (2,2), (2,1,1)$ and $(1,1,1,1)$.
It is important to notice that equation (\ref{kesimaderparti}) provides all the possible integer
partitions of $k$, in a recursive way.

Let us calculate the limits of the derivatives of $f$ when $x$ decreases to zero. For the first derivative,
using l'H\^{o}pital three times, we have
\begin{eqnarray}\label{prima}
\lim_{x\downarrow 0} f^{(1)}(x) & = & \lim_{x\downarrow 0} \frac{2\exp(-1/x^2)}{x^3} =  
\lim_{t \uparrow\infty}\frac{2\exp(-1/(1/t)^2)}{(1/t)^3} \nonumber\\
& = & \lim_{t\uparrow\infty}\frac{12t}{(2+(2t)^2)\exp(t^2)} = 
\lim_{t\uparrow\infty}\frac{12}{(12t+(2t)^3)\exp(t^2)}=0
\end{eqnarray}
Similarly, for the second derivative using L'H\^{o}pital six times we get
\begin{equation}\label{biprima}
\lim_{x\downarrow 0} f^{(2)}(x)= \lim_{t \uparrow\infty} \frac{5! 4!}{(5!+ 6!t^2 +4 5!t^4+64t^6)\exp(t^2)}=0.
\end{equation}
In general, as in equations (\ref{prima}) and (\ref{biprima}), we have
$$ \lim_{x \downarrow 0} f^{(k)}(x)=\lim_{t \uparrow \infty} \frac{Q_{3k}(t)}{\exp(t^2)}=0,$$
where $Q_{3k}$ is a polynomial of degree $3k$, and  using l'H\^{o}pital $3k$  times we obtain the
limit of a quotient of a constant divided by a polynomial of order $3k$ times $\exp(t^2)$, which goes to zero
as $t$ increases to infinity. Therefore, if we define 
$f^{(k)}(0)=\lim_{x\downarrow 0} f^{(k)}(x)=0$ for every $k\geq 0$, we have that the Taylor (McLaurin) power series 
of $f$ around $x_0=0$ is given by
\begin{equation}\label{uselesstaylor}
f(x)=\exp(-1/x^2)=\sum_{k=0}^{\infty} \frac{f^{(k)}(0)}{k!} x^k=0\quad\mbox{for every}\quad x\in\mathbb{R}.
\end{equation}
Of course, equation (\ref{uselesstaylor}) is false for every $x\not = 0$, hence, it is useless. However, when we use points $x_0$ which are not too close to zero, the Taylor expansion around $x_0$ is a good approximation of $f$ around $x_0$. We used two values  $x_0=1$ and $x_0=2$, in order to obtain the $m^{\mbox{th}}$ Taylor polynomial approximation of $f(x)=\exp(-1/x^2)$, we used values of  $m=5, 20, 50, 75$ and $m=100$.

In Table 2 we give the approximated values of $f(x)$ when $x_0=1$, and in Table 3 we give the approximated
values of $f(x)$ when $x_0=2$ for the different values of $m$ given above.

\begin{table}[t!]
\caption{Approx. Values of $f(x)=\exp(-1/x^2)$ when  $x_0=1$ and for some values of $x$ and $m$ varying between $m=5$ and $m=100$}
\renewcommand{\arraystretch}{1}
\begin{tabular}{|c|c|c|c|c|c|c|}\hline
$\mbox{Value of}\,x$ & $m=5$ & $m=20$ & $m=50$ & $m=75$ & $m=100$ &  $\mbox{Value of}\,\, f(x)$ \\ \hline
0.01& 1.23881 &19.6154 &3070.02 & 129968 &2.61 $\times 10^{6}$ &{\bf 0.0} \\ \hline
0.1 & 0.70322 &3.78329 &18.4717 &106.978  & 208.791 &{\bf 3.72 $\times 10^{-44}$}  \\ \hline
0.2 & 0.32537 &0.43042 &0.01066 & 0.011824 & 0.001220 &{\bf 1.38 $\times 10^{-11}$}  \\ \hline
0.4 & 0.02845 &0.00300 &{\bf 0.0019304}&{\bf 0.0019304} &{\bf 0.0019304}  &{\bf 0.0019304} \\ \hline
0.6 & 0.06111 &{\bf 0.062176} &{\bf 0.062176} &{\bf 0.062176} &{\bf 0.062176}  &{\bf 0.062176} \\ \hline
0.8 & 0.20957 &{\bf 0.209611} &{\bf 0.209611} &{\bf 0.209611} &{\bf 0.209611}  &{\bf 0.209611} \\ \hline
1  & {\bf 0.367879}&{\bf 0.367879} &{\bf 0.367879}    &{\bf 0.367879}    &{\bf 0.367879}  &{\bf 0.367879} \\ \hline
1.2& 0.49930 &{\bf 0.499352} &{\bf 0.499352} &{\bf 0.499352} &{\bf 0.499352} &{\bf 0.499352} \\ \hline
1.4& 0.59773 &{\bf 0.600373}&{\bf 0.600373}&{\bf 0.600373}  &{\bf 0.600373} &{\bf 0.600373} \\ \hline
1.6& 0.64903 &0.676646 &{\bf 0.676634}   &{\bf 0.676634} &{\bf 0.676634}  &{\bf 0.676634} \\ \hline
1.8& 0.59246 &0.676405 & 0.750045  & 0.733865 & 0.73447  &{\bf 0.734444} \\ \hline
1.9& 0.48249 &0.121906 &6.75519 &-3.57417  & 4.58669  &{\bf 0.758048} \\ \hline
2  & 0.28204 &-4.6187 &1227.56 &-12604    & 158190  &{\bf 0.778801} \\ \hline
\end{tabular}
\end{table}

\begin{table}[t!]
\caption{Approx. Values of $f(x)=\exp(-1/x^2)$ when  $x_0=2$ and for some values of $x$ and $m$ varying between $m=5$ and $m=100$}
\renewcommand{\arraystretch}{1}
\begin{tabular}{|c|c|c|c|c|c|c|}\hline
$\mbox{Value of}\,x$ & $m=5$ & $m=20$ & $m=50$ & $m=75$ & $m=100$ &  $\mbox{Value of}\,\, f(x)$ \\ \hline
0.01& -1.29843 &-0.52540 &-75.4647 & 1644.49 & -6243.53 &{\bf 0.0} \\ \hline
0.2 & -0.79126 &0.36364 &0.117268 &0.883608  & -0.50432 &{\bf 1.38 $\times 10^{-11}$}  \\ \hline
0.4 & -0.37613 &0.07556 &0.003711 & 0.001986 & 0.0019276 &{\bf 0.0019304}  \\ \hline
0.8 & -0.18205 &0.020977 &{\bf 0.209611}&{\bf 0.209611} &{\bf 0.209611}  &{\bf 0.209611} \\ \hline
1.2 & 0.49881 &{\bf 0.499352} &{\bf 0.499352} &{\bf 0.499352} &{\bf 0.499352}  &{\bf 0.499352} \\ \hline
1.6 & 0.676636 &{\bf 0.676634} &{\bf 0.676634} &{\bf 0.676634} &{\bf 0.676634}  &{\bf 0.676634} \\ \hline
2  & {\bf 0.778801}&{\bf 0.778801} &{\bf 0.778801}    &{\bf 0.778801}    &{\bf 0.778801}  &{\bf 0.778801} \\ \hline
2.4& 0.84063 &{\bf 0.840624} &{\bf 0.840624} &{\bf 0.840624} &{\bf 0.840624} &{\bf 0.840624} \\ \hline
2.8& 0.88070 &{\bf 0.88024}&{\bf 0.88024}&{\bf 0.88024}  &{\bf 0.88024} &{\bf 0.88024} \\ \hline
3.2& 0.91206 &0.906923 &{\bf 0.906961}   &{\bf 0.906961} &{\bf 0.906961}  &{\bf 0.906961} \\ \hline
3.6& 0.95348 &0.911864 & 0.925403  & 0.733865 & {\bf 0.925741}  &{\bf 0.925741} \\ \hline
3.8& 0.98814 &0.778038&0.80337 &-3.57417  & 0.954813  &{\bf 0.933091} \\ \hline
4  & 1.04063 &-0.39957 &-25.56 &-63.9548    & 856.815  &{\bf 0.939413} \\ \hline
\end{tabular}
\end{table}

\begin{table}[t!]
\caption{Values of $|c_n|^{1/n}$ for some values of $n$ in the Taylor expansion at $x_0=1$}
\renewcommand{\arraystretch}{1}
\begin{tabular}{|c|c|c|c|c|c|c|c|c|}\hline
$n$ & 10 & 50 & 100 & 500 & 750 & 1000 & 1500 & 2000 \\ \hline
$|c_n|^{1/n}$ &1.1044 &1.1678 &1.1405 &1.1069 &1.0962 &1.0865 &1.0785 &1.0714  \\ \hline
\end{tabular}
\end{table}

\begin{table}[t!]
\caption{Values of $|c_n|^{1/n}$ for some values of $n$ in the Taylor expansion at $x_0=2$}
\renewcommand{\arraystretch}{1}
\begin{tabular}{|c|c|c|c|c|c|c|c|c|}\hline
$n$ & 10 & 50 & 100 & 500 & 750 & 1000 & 1500 & 2000 \\ \hline
$|c_n|^{1/n}$ &0.4941 &0.5407 &0.5377 &0.5298 &0.5290 &0.5268 &0.5239 &0.5219 \\ \hline
\end{tabular}
\end{table}

In Tables 2 and 3 we observe that depending on the value of $x_0$ in the Taylor expansion of $f$, the radius of convergence changes accordingly. In fact, in Tables 4 and 5 we observe that using equation (\ref{radius}), the values of $|c_m|^{1/m}$ for $x_0=1$ are a little smaller than one, and in the case of $x_0=2$
the values of $|c_m|^{1/m}$ are a little smaller than two. This behavior may be understood, because as  we observed 
for $|x|\leq 1/27.29712=0.0366339$ the value of $f(x)$ is practically zero.

If we obtain the Taylor expansion around $x_0=0$ of the function $g(x)=\ln(a+x)$ for any $a>0$, it is easy to see
that $g^{(k)}(x)=\frac{(-1)^{k+1}(k-1)!}{(a+x)^k}$ for any $k\geq 1$.
So, the coefficients of $x^k$ in the Taylor expansion are $c_k=\frac{g^{(k)}(0)}{k!}=\frac{(-1)^{k+1}}{k a^k}$.
Hence, $|c_k|^{1/k}=\frac{1}{a} \left(\frac{1}{k}\right)^{1/k}$ for every $k\geq 1$. Therefore,
using the formula (\ref{radius}) for the radius of convergence of the Taylor series we get
that
$$ \frac{1}{r}=\limsup_{k\rightarrow\infty}  \frac{1}{a} \left(\frac{1}{k}\right)^{1/k}=\frac{1}{a}.$$
So, the radius of convergence is given by $r=a$.  It must be observed that for any $a>0$, the convergence of the
Taylor series fails not only for $x=-a$ when $\ln(a+(-a))=\ln(0)=-\infty$, but also fails when
$x=a$ even when $\ln(a+a)$ is perfectly defined. This phenomenon occurs also for the Taylor expansions
of $f(x)=\exp(-1/x^2)$ in Tables 2 and 3.

\vskip0.1cm
\noindent {\bf 3 Solving differential equations by differentiation.}
\vskip0.1cm

\noindent When we solve a differential equation it is of particular interest to find solutions $f$ which are infinitely differentiable around a point $x_0\in \mathbb{R}$, because in this case the Taylor expansion of
$f$ around $x_0$ is a very nice solution. As we saw in the previous sections in some instances it is possible to find the  solution using some Taylor series of very well known functions. And even in the case that we do not know which function is represented by the Taylor series, we can provide a very close approximation to the solution (e.g. \cite{B}), using that the computer's packages nowadays provide complex recursive equations that can be solved in very short times .  In fact, in many instances, it is possible to obtain the exact form of the constants needed in the Taylor expansion around a point $x_0\in\mathbb{R}$, using a recursive formula which is easy to be evaluated, so, we can write the exact form of the Taylor expansion of the solution.

We observed in the previous sections that it is possible to provide iterative formulae to evaluate the derivatives
of the function, which are based on the initial values of the solution function $f$. Therefore, we want
to find an adequate set of minimal conditions on the differential equation that may provide
recursive equations to find all the derivatives of $f$ at $x_0$. Note that the values of $ \{f^{(k)}(x_0)\}_{k=0}^{\infty}=\{b_k\}_{k=0}^{\infty}$
are required to find  the radius of convergence for  the Taylor series, using results like the Cauchy-Hadamard theorem.

A new theoretical result  in this direction is to solve linear differential equations of the form:

\begin{equation}\label{fnsum}
f^{(0)}(x)= g_1^{(0)}(x) f^{(1)}(x)+g_2^{(0)}(x) f^{(2)}(x) +\cdots +g_n^{(0)}(x) f^{(n)}(x),
\end{equation}
where $n\in\mathbb{N}$, $g_1^{(0)}, g_2^{(0)},\ldots ,g_n^{(0)}$ are infinitely differentiable functions 
defined in an open set $(a,b)$ where
$-\infty \leq a<b\leq \infty$. Let $x_0\in (a,b)$ and assume the initial conditions
$f^{(i)}(x_0)=b_i\in \mathbb{R}$  for $i\in \{1,\ldots ,n\}$, and $g_n^{(0)}(x_0)\not= 0$.

\noindent {\bf Theorem 1.}
{\it The solution of equation (\ref{fnsum}) is known when we have all the values of $b_{i},$  which are
the coefficients in its Taylor series.  We only need to define $b_{n+m},$ for every $m \geq 1$ by
\begin{equation}\label{fn+m1}
b_{n+m} = \frac{1}{g_n^{(0)}(x_0)}\left[ b_m -\sum_{j=1}^{n-1} b_{j+m} g_j^{(0)}(x_0) -\sum_{k=1}^{m}
\left(\begin{array}{c}
m 
\\ 
k
\end{array}\right) \sum_{j=1}^{n} b_{j+(m-k)} g_j^{(k)}(x_0)\right]
\end{equation}}

\noindent  {\bf Proof:}  First observe that $b_0=f(x_0)=f^{(0)}(x_0)=\sum_{j=1}^{n} b_j g_j^{(0)}(x_0)$.
Now derivating (\ref{fnsum}) with respect to $x$ we obtain

\begin{equation}\label{fn1}
f^{(1)}(x)=\sum_{j=1}^{n} g_j^{(0)}(x) f^{(j+1)}(x) +\sum_{k=1}^{n} g_k^{(1)}(x) f^{(k)}(x)
\end{equation}
Evaluating (\ref{fn1}) at $x=x_0$ and using that $g_n^{(0)}(x_0)\not = 0$, we get
\begin{eqnarray}\label{fn1at0}
b_{n+1}=f^{(n+1)}(x_0)& = & \frac{1}{g_n^{(0)}(x_0)}
\left[b_1-b_2g_1^{(0)}(x_0)-b_3 g_2^{(0)}(x_0)-\cdots -b_ng_{n-1}^{(0)}(x_0)\right.\nonumber \\
& & \left. -b_1g_1^{(1)}(x_0)-b_2g_2^{(1)}(x_0)-\cdots - b_ng_n^{(1)}(x_0)\right]\nonumber\\
& = & \frac{1}{g_n^{(0)}(x_0)}\left[ b_1[1-g_1^{(1)}(x_0)]-
\sum_{j=2}^{n} b_j[g_{j-1}^{(0)}(x_0)+g_j^{(1)}(x_0)]\right]\nonumber\\
& = & \frac{1}{g_n^{(0)}(x_0)}\left[ -\sum_{j=1}^{n} b_j\left(g_{j-1}^{(0)}(x_0)+g_j^{(1)}(x_0)\right) \right],
\end{eqnarray}
where $g_0^{(0)}(x)=-1$ for every $x\in (a,b)$. Of course $b_{n+1}$ and $b_0$ only depend on  
$b_i$ for $i\in \{1,2,\ldots ,n\}$ the initial values. Differentiating (\ref{fn1}) we obtain

\begin{eqnarray}\label{fn2}
f^{(2)}(x)& =&\sum_{j=1}^{n} g_j^{(0)}(x) f^{(j+2)}(x) +\sum_{k=1}^{n} g_k^{(1)}(x) f^{(k+1)}(x)\nonumber\\
& & +\sum_{i=1}^{n} g_i^{(1)}(x) f^{(i+1)}(x) +\sum_{l=1}^{n} g_l^{(2)}(x)f^{(l)}(x)\nonumber\\
& = & \sum_{j=1}^{n} g_j^{(0)}(x) f^{(j+2)}(x) +2\sum_{k=1}^{n} g_k^{(1)}(x)f^{(k+1)}(x)+
\sum_{l=1}^{n} g_l^{(2)}(x) f^{(l)}(x).
\end{eqnarray}
Evaluating (\ref{fn2}) at $x=x_0$, using the initial conditions and the fact that $g_0^{(0)}(x)=-1$, we get

\begin{eqnarray}\label{fn2at0}
b_2=f^{(2)}(x_0) & = & \sum_{j=1}^{n-1} g_j^{(0)}(x_0) f^{(j+2)}(x_0) + g_n^{(0)}(x_0) f^{(n+2)}(x_0) \nonumber\\
& & + 2\sum_{k=1}^{n} g_k^{(1)}(x_0) f^{(k+1)}(x_0) +\sum_{l=1}^{n} g_l^{(2)}(x_0) f^{(l)}(x_0)\nonumber\\
& = & \sum_{j=1}^{n-1} b_{j+2} g_j^{(0)}(x_0)+ g_{n}^{(0)}(x_0) f^{(n+2)}(x_0)\nonumber\\
& & +2\sum_{k=1}^{n} b_{k+1}g_k^{(1)}(x_0) +\sum_{l=1}^{n} b_l g_l^{(2)}(x_0).
\end{eqnarray}
Solving equation (\ref{fn2at0}) for $f^{(n+2)}(x_0)$ and using that $g_n^{(0)}(x_0)\not = 0$, we obtain that
\begin{eqnarray}\label{bn+2}
b_{n+2}:=f^{(n+2)}(x_0) & = & \frac{1}{g_n^{(0)}(x_0) }\left[ b_2 - \sum_{j=1}^{n-1} b_{j+2}g_j^{(0)}(x_0)\right. \nonumber\\
& &\left. - 2\sum_{k=1}^{n} b_{k+1}g_k^{(1)}(x_0)-\sum_{l=1}^{n} b_l g_l^{(2)}(x_0)\right] \nonumber\\
& = & \frac{1}{g_n^{(0)}(x_0)}\left[b_2-(b_3g_1^{(0)}(x_0)+b_4 g_2^{(0)} (x_0)+\cdots+ b_{n+1}g_{n-1}^{(0)}(x_0)) \right.\nonumber\\
& & -2(b_2g_1^{(1)}(x_0)+b_3 g_2^{(1)}(x_0)+\cdots + b_{n+1}g_n^{(1)}(x_0)) \nonumber\\
& & \left. -(b_1g_1^{(2)}(x_0)+b_2 g_2^{(2)}(x_0)+\cdots +b_n g_n^{(2)}(x_0))\right]\nonumber\\
& = & \frac{1}{g_n^{(0)}(x_0)}\left[-b_1g_1^{(2)}(x_0)\right. -\sum_{i=2}^{n} b_i\left\{ g_{i-2}^{(0)}(x_0)+
2g_{i-1}^{(1)}(x_0)+g_i^{(2)}(x_0)\right\}\nonumber\\
& &\left. -b_{n+1}\left\{ g_{n-1}^{(0)}(x_0)+ 2 g_n^{(1)}(x_0)\right\} \right]
\end{eqnarray}
Again, using equation (\ref{fn1at0}) and (\ref{bn+2}) we get that $b_{n+2}$ is a function of
$b_i$ for $i\in \{1,2,\ldots ,n\}$.

\noindent In general, it is easy to see that for every $m\geq 0$ we have that 

\begin{equation}\label{fm}
f^{(m)}(x)=\sum_{k=0}^{m} 
\left(\begin{array}{c}
m 
\\ 
k
\end{array}\right)\sum_{j=1}^{n} g_j^{(k)}(x) \cdot f^{(j+(m-k))}(x)
\end{equation}
for every $x\in (a,b)$. By induction let us assume that we know that  $b_k=f^{(k)}(x_0)$ is a function of $b_1,b_2,\ldots b_n$
for every $0\leq k\leq n+m-1$, let us prove that $b_{n+m}=f^{(n+m)}(x_0)$ is also a function
of $b_1,\ldots ,b_n$. First we observe that from equation (\ref{fm})

\begin{eqnarray}\label{fmat0}
b_m=f^{(m)}(x_0) & = & \left(\begin{array}{c}
m 
\\ 
0
\end{array}\right)\sum_{j=1}^{n-1} g_j^{(0)}(x_0)\cdot f^{(j+m)}(x_0)+
\left(\begin{array}{c}
m 
\\ 
0
\end{array}\right) g_n^{(0)}(x_0)\cdot f^{(n+m)}(x_0)\nonumber\\
& & +\sum_{k=1}^{m}\left(\begin{array}{c}
m 
\\ 
k
\end{array}\right)\sum_{j=1}^{n} g_j^{(k)}(x_0)\cdot f^{(j+(m-k))}(x_0).
\end{eqnarray}
We  want to solve (\ref{fmat0}) for  $b_{n+m}= f^{(n+m)}(x_0)$. Therefore,

\begin{eqnarray}\label{fn+m}
b_{n+m} &=& f_{(n+m)}(x_0)  =  \frac{1}{g_n^{(0)}(x_0)}\left[ b_m -\sum_{j=1}^{n-1} g_j^{(0)}(x_0)\cdot f^{(j+m)}(x_0)\right.\nonumber\\
 &-&   \sum_{k=1}^{m} \left(\begin{array}{c}
m 
\\ 
k
\end{array}\right) \left. \sum_{j=1}^{n} g_j^{(k)}(x_0)\cdot f^{(j+(m-k)}(x_0)\right] \nonumber\\
& = &\frac{1}{g_n^{(0)}(x_0)}\left[ b_m -\sum_{j=1}^{n-1} b_{j+m} g_j^{(0)}(x_0) -\sum_{k=1}^{m}
\left(\begin{array}{c}
m 
\\ 
k
\end{array}\right) \sum_{j=1}^{n} b_{j+(m-k)} g_j^{(k)}(x_0)\right]
\end{eqnarray}
This finishes the proof.  \qed

Observe that in equation (\ref{fn+m}), the largest index for the $b's$ is $n+(m-1)$ which can be found
in the middle term when $j=n-1$ and in the third term when $j=n$ and $k=1$, since $b_{j+(m-k)}=b_{n+(m=1)}$. Besides, the 
smallest index appears in the third term when $j=1$ and $k=m$, where $b_{j+(m-k)}=b_1$. Therefore, using the induction 
hypothesis we obtain that $b_{n+m}$ only depends on the initial conditions $b_i$ for $i\in\{ 1,2,\ldots, n\}$.

\setcounter{equation}{33}
\vskip0.1cm
\noindent {\bf 4 Bivariate Functions.}
\vskip0.1cm
\vskip0.1cm
\noindent {\bf 4.1 Bivariate Taylor Series  and Differential Equations.}
\vskip0.1cm

\noindent Let $f:\mathbb{R}^2\rightarrow \mathbb{R}$ be an infinitely differentiable function, that is,
using the  standard notation for partial derivatives
\begin{equation}\label{parciales}
\frac{\partial^{j+k}}{\partial x^{j} \,\partial y^{k}} f(x,y)\quad\mbox{exists for every}\quad
(x,y)\in\mathbb{R}^2 \,\,\mbox{and for every}\,\,j,k\in\{0,1,2,\ldots\}.
\end{equation}
From (\ref{parciales}),  if $j=1$ and $k=0$ we will simply write $\frac{\partial}{\partial x}f(x,y)$ instead of
$\frac{\partial^{1+0}}{\partial x^{1}\, \partial y^{0}}f(x,y)$, similarly for $j=0$ and $k=1$.
Also, if $j=k=1$ we will write $\frac{\partial^{2}}{\partial x \partial y}f(x,y)$  instead of
$\frac{\partial^{2}}{\partial x^{1} \partial y^{1}}f(x,y)$. In fact,
$\frac{\partial^{2}}{\partial x \partial y}f(x,y)=\frac{\partial}{\partial x}\left(\frac{\partial}{\partial y}f(x,y)\right)$
and
$\frac{\partial^{2}}{\partial y \partial x}f(x,y)=\frac{\partial}{\partial y}\left(\frac{\partial}{\partial x}f(x,y)\right)$,
and if these partial derivatives are continuous by Clairaut's theorem
$\frac{\partial^{2}}{\partial x \partial y}f(x,y)=\frac{\partial^{2}}{\partial y \partial x}f(x,y)$, that is,
the partial derivatives can be exchanged, this can be easily generalized to partials of higher orders.
Besides, $f$ has a Taylor (McLaurin) expansion in double series around $(x_0,y_0)=(0,0)$ given by:
\begin{eqnarray}\label{taylorbi}
f(x,y)&=&\sum_{j=0}^{\infty}\sum_{k=0}^{\infty} c_{j\, k} x^j y^k\nonumber\\
& = & c_{0 0} +c_{1 0} x+  c_{0 1} y +c_{2 0} x^2+ c_{1 1} xy+ c_{0 2} y^2+\cdots ,
\end{eqnarray}
where $c_{j k}= \frac{\partial^{j+k}}{\partial x^{j} \,\partial y^{k}} f(0,0)/(j! \, k!)$
for every $j, k$ non negative integers. Of course, 
$\frac{\partial^{0}}{\partial x^{0} \,\partial y^{0}} f(x,y)=f(x,y)$ for every $(x,y)\in\mathbb{R}^2$.
The Taylor expansion around an arbitrary point $(x_0,y_0)\in\mathbb{R}^2$, is given basically by
equation (\ref{taylorbi}) with obvious changes, see equation (\ref{taylor}).
\vskip .3cm
\noindent {\bf Example 4.1.1} Let $f:\mathbb{R}^2\rightarrow \mathbb{R}$ be an infinitely differentiable function that satisfies the
partial differential equation
\begin{equation}\label{trivial}
\frac{\partial}{\partial x}f(x,y)=\frac{\partial}{\partial y}f(x,y)=f(x,y)
\quad\mbox{for every}\quad (x,y)\in\mathbb{R}^2,
\end{equation}
with initial condition $f(0,0)=1$.  To solve equation (\ref{trivial}) we first note that
$$ \frac{\partial^{j+k}}{\partial x^{j} \,\partial y^{k}}f(x,y)=f(x,y)
\quad\mbox{for every} \quad (x,y)\in \mathbb{R}^2 \,\,\mbox{and for every}\,\, j,k\geq 0.$$
From the initial condition, we have $\frac{\partial^{j+k}}{\partial x^{j} \,\partial y^{k}} f(0,0)=f(0,0)=1,$
for every $j,k$ nonnegative integers. So, using equation (\ref{taylorbi}) we obtain 
\begin{eqnarray}\label{solutionpartial1}
f(x,y)&=&\sum_{j=0}^{\infty}\sum_{k=0}^{\infty} \frac{\frac{\partial^{j+k}}{\partial x^{j} \,\partial y^{k}}f(0,0)}{j! \, k!}
           x^j y^k \nonumber\\
& = & \sum_{j=0}^{\infty}\sum_{k=0}^{\infty}\frac{1}{j! \, k!} x^j  y^k\nonumber\\
& = & \sum_{j=0}^{\infty}\frac{x^j}{j!} \cdot\sum_{k=0}^{\infty} \frac{y^k}{k!}\nonumber\\
& = & \exp(x) \cdot \exp(y)\nonumber\\
& = & \exp(x+y),
\end{eqnarray}
for every $(x,y)\in\mathbb{R}^2$. It is clear that $f$ given in equation (\ref{solutionpartial1}) is a solution
for equation (\ref{trivial}), with initial condition $f(0,0)=1$.
\vskip .3cm
\noindent {\bf Example 4.1.2} As a second example, let $f:\mathbb{R}^2\rightarrow \mathbb{R}$ be an infinitely differentiable function that satisfies
the partial differential equation
\begin{equation}\label{notsotrivial}
\frac{\partial}{\partial x} f(x,y)=y f(x,y) \quad\mbox{and}\quad \frac{\partial}{\partial y} f(x,y)=x f(x,y),
\end{equation}
with initial condition $f(0,0)=1$. From equation (\ref{notsotrivial}) we have that
$\frac{\partial}{\partial x} f(0,0)=0$ and $\frac{\partial}{\partial y} f(0,0)=0$,
we also have
$$\frac{\partial^2}{\partial x \partial y} f(x,y)= \frac{\partial}{\partial x} (x \cdot f(x,y))
= f(x,y) +x\frac{\partial}{\partial x}f(x,y)=f(x,y)+x\cdot y f(x,y).$$
So, $\frac{\partial^2}{\partial x \partial y} f(0,0)=1$. From equation (\ref{notsotrivial}), we also have 
$$\frac{\partial^2}{\partial x^2}f(x,y)=\frac{\partial}{\partial x}(y f(x,y))=y\frac{\partial}{\partial x}f(x,y)
=y^2 f(x,y).$$
Hence, $\frac{\partial^2}{\partial x^2}f(0,0)=0$. Analogously, $\frac{\partial^2}{\partial y^2}f(x,y)=x^2 f(x,y)$
and $\frac{\partial^2}{\partial y^2}f(0,0)=0$. In general,
$$\frac{\partial^k}{\partial x^k}f(x,y)=y^k f(x,y) \quad\mbox{and}\quad 
\frac{\partial^k}{\partial y^k}f(x,y)=x^k f(x,y)\,\,\,\mbox{for every}\,\,\, k\geq 1.$$
So, $\frac{\partial^k}{\partial x^k}f(0,0)=0=\frac{\partial^k}{\partial y^k}f(0,0)$ for every $k\geq 1$. Now
$$ \frac{\partial^3}{\partial y\, \partial x^2}f(x,y)=\frac{\partial}{\partial y}(y^2 f(x,y))=
2y f(x,y)+ y^2 \frac{\partial}{\partial y}f(x,y)
=(2y+x y^2)f(x,y).$$
So, $\frac{\partial^3}{\partial y\, \partial x^2}f(0,0)=0$. Analogously,
$\frac{\partial^3}{\partial y^2\, \partial x}f(x,y)=(2x+x^2 y)f(x,y)$ and 
$\frac{\partial^3}{\partial y^2\, \partial x}f(0,0)=0$. Now we observe that

$$ \frac{\partial^4}{\partial y^2\, \partial x^2}f(x,y)=\frac{\partial}{\partial y}((2y+x y^2)f(x,y))
=(2+4xy+x^2y^2)f(x,y).$$
Hence, $\frac{\partial^4}{\partial y^2\, \partial x^2}f(0,0)=2=2!$. It is not difficult to see that
$\frac{\partial^5}{\partial y^2\, \partial x^3}f(x,y)=(6y+6x y^2+x^2 y^3)f(x,y)$ and $\frac{\partial^ 6}{\partial y^3\, \partial x^3}f(x,y)=(6+18xy+9x^2y^2+x^3y^3)f(x,y).$
So, $\frac{\partial^5}{\partial y^2\, \partial x^3}f(0,0)=0$ and 
$\frac{\partial^6}{\partial y^3\, \partial x^3}f(0,0)=6=3!$. In general we observe that
\begin{equation}\label{todas}
\frac{\partial^{j+k}}{\partial y^k \,\partial x^j}f(0,0)=\left\{ \begin{array}{lcl}
j! & \quad\mbox{if}\quad & j=k\nonumber\\
0  & \quad\mbox{if}\quad & j\not=k.
\end{array}\right.
\end{equation}
Therefore, using equation (\ref{taylorbi}) we have that
\begin{equation}\label{solutionpartial2}
f(x,y)=\sum_{j=0}^{\infty}\frac{j!}{j! \, j!} x^j y^j=\sum_{j=0}^{\infty}\frac{(xy)^j}{j!}=\exp(xy).
\end{equation}
It is trivial to see that equation (\ref{solutionpartial2}) is a solution of
equation (\ref{notsotrivial}) with  initial condition $f(0,0)=1$.
\vskip .3cm
\noindent {\bf Example 4.1.3} As a final example we take $f:\Omega\rightarrow\mathbb{R}$ where $\Omega\subset \mathbb{R}^2$ is an open connected set to be defined,
such that $(0,0)\in\Omega$,
and we assume that $f$ is infinitely differentiable in $\Omega$. Assume that $f$ satisfies the 
partial differential equations:
\begin{equation}\label{powerseries}
\frac{\partial}{\partial x} f(x,y)=\frac{\partial}{\partial y} f(x,y)= f^2(x,y), \quad\mbox{for every}\quad
(x,y)\in\Omega,
\end{equation}
with initial condition $f(0,0)=1$. Then from (\ref{powerseries}) we have $\frac{\partial}{\partial x} f(0,0)=1$
and $\frac{\partial}{\partial y} f(0,0)=1$. Differentiating (\ref{powerseries}) we get
\begin{equation}\label{parttwo}
\frac{\partial^{2}}{\partial y^{2}} f(x,y)=\frac{\partial^{2}}{\partial x\,\partial y} f(x,y)=
\frac{\partial^{2}}{\partial x^{2}} f(x,y)=2f(x,y)\frac{\partial}{\partial x} f(x,y)=2 f^3(x,y),
\end{equation}
for every $(x,y)\in\Omega$. So, $\frac{\partial^{2}}{\partial y^{2}} f(0,0)=
\frac{\partial^{2}}{\partial x\,\partial y} f(0,0)=\frac{\partial^{2}}{\partial x^{2}} f(0,0)=2$.
From equation (\ref{parttwo}) it is clear that
$$\frac{\partial^{3}}{\partial y^{3}} f(x,y)=\frac{\partial^{3}}{\partial x\,\partial y^{2}} f(x,y)=
\frac{\partial^{3}}{\partial x^{2}\,\partial y} f(x,y)=\frac{\partial^{3}}{\partial x^{3}} f(x,y)=3! f^4(x,y).
$$
Therefore, it is easy to see that
$$
\frac{\partial^{j+k}}{\partial y^{k}\, \partial x^{j}}f(0,0)=(j+k)!
\,\,\mbox{for every}\,\, j,k \,\, \mbox{nonegative integers}.
$$
Hence, using the equation of Taylor (McLaurin) expansion (\ref{taylorbi}), assuming that
$|x+y|<1$ and   rearranging terms of the double sum 
\begin{eqnarray}\label{solutionpartial3}
f(x,y) & = & \hspace{-5pt}\sum_{j=0}^{\infty}\sum_{k=0}^{\infty} \frac{(j+k)!}{j!\, k!} x^{j} y^{k} 
 =  \sum_{j=0}^{\infty}\sum_{k=0}^{\infty}\left(\begin{array}{c}
       j+k \\
			  j
			\end{array}\right) x^{j} y^{k}\nonumber\\
& = & \left(\begin{array}{c}
       0 \\
			  0
			\end{array}\right) x^{0}y^{0} +\left[\left(\begin{array}{c}
                                      1 \\
			                                0
			                                     \end{array}\right) x^{0}y^{1}+
																					  \left(\begin{array}{c}
                                      1 \\
			                                1
			                                     \end{array}\right) x^{1}y^{0}\right] +\cdots\nonumber	\\
& = &	\sum_{n=0}^{\infty}\sum_{k=0}^{n}		\left(\begin{array}{c}
                                      n \\
			                                k
			                                     \end{array}\right)		x^{k} y^{n-k}	
 =  \sum_{n=0}^{\infty} (x+y)^n \nonumber\\
& = & \frac{1}{(1-x-y)},
\end{eqnarray}
the last equality in (\ref{solutionpartial3}) follows from hypothesis, since  we have a convergent power series.
Clearly $\Omega = \{ (x,y) \in  \mathbb{R}^{2} :  |x+y| < 1 \}.$
It is easy to see that the function $f(x,y)=1/(1-x-y)$ is a solution of the partial differential equation
(\ref{powerseries}), with initial condition $f(0,0)=1$. It is also clear that $f$ is not well defined
if $x+y=1$ which is straight line with slope $m=-1$. 

This example can be easily generalized to any dimension $n$ by taking $f:\Omega_n\rightarrow \mathbb{R}$ where
$\Omega_n\subset\mathbb{R}^n$ is an open connected set, 
such that $(0,\ldots ,0)\in\Omega$, and $f$ is infinitely differentiable  in $\Omega_n$.
Assume that $f$ satisfies the partial differential equation
$$
\frac{\partial}{\partial x_i} f(x_1,\ldots ,x_i,\ldots ,x_n)
= f^2(x_1,\ldots ,x_i,\ldots , x_n)\quad\mbox{for every}\,\,
(x_1,\ldots ,x_n)\in\Omega_n,
$$
and for every $i\in\{1,2,\ldots ,n\}$, with initial condition $f(0,\ldots ,0)=1$. Reasoning exactly as above, assuming that 
$|x_1+\cdots +x_n|<1$ and   rearranging terms of the multiple sum we obtain that
$$ f(x_1,\ldots ,x_n)=\frac{1}{(1-x_1-x_2-\cdots -x_n)}\quad\mbox{for every}\quad
(x_1,\ldots , x_n)\in\Omega_n.$$
In the theory of multivariate power series we have the following notation
$$ g({\bf x})=g(x_1,\ldots ,x_n)=\sum_{\alpha} a_{\alpha} {\bf x}^{\alpha},$$
where $\alpha =(\alpha_1,\ldots \alpha_n)\in\{0,1,2,\ldots\}^{n}$ is any vector with nonnegative integer coordinates,
$a_{\alpha}$ are real or complex constants and ${\bf x}^{\alpha}=
x_{1}^{\alpha_1}\cdot x_{2}^{\alpha_2}\ldots \cdot x_{n}^{\alpha_n}$.
Let us denote by $|\alpha|=\sum_{i=1}^{n} \alpha_i$, then $|\alpha|$ is always a nonnegative integer, and we call it
the {\bf size} of $\alpha$. Of course, all this notation can be used for the Taylor multivariate series expansions
around ${\bf x}_0=(0,\ldots ,0)$, in this case if $\alpha=(\alpha_1,\ldots \alpha_n)$, then
$$ C_{\alpha}=\frac{\frac{\partial^{|\alpha|}}{\partial x_{1}^{\alpha_1} \,\partial x_{2}^{\alpha_2}\cdots \partial x_{n}^{\alpha_n}}
f(0,0,\ldots ,0)}{ (\alpha_1)! \cdot (\alpha_2)!\cdots (\alpha_{n})!}.$$
Using a generalized version of (\ref{radius}),  we may write using \cite{FJ}
\begin{equation}\label{buena?}
\frac{1}{R} =\limsup_{m\rightarrow\infty} 
\left[ \sup_{\{\alpha\in\{0,1,2,\ldots\}^{n} \,|\,\,\, |\alpha|=m\}}| C_{\alpha}|^{1/|\alpha|}\right].
\end{equation}
In our case it is easy to see that the supremum in equation (\ref{buena?}) is attained when
$\alpha_1=\alpha_2=\cdots=\alpha_n$ for $m=n*k$ and $k\geq 1$  an integer.
Let $|d_{\alpha_1,\ldots ,\alpha_n}|^{1/(\alpha_1*n)}
=\sup_{\{\alpha\in\{0,1,2,\ldots\}^{n} \,|\,\,\, |\alpha|=n*\alpha_1\}}| C_{\alpha}|^{1/|\alpha|}$,
where $\alpha_1=\alpha_2=\cdots =\alpha_n=k$.
In Table 6 we obtain the results obtained for different values of $k$ in dimensions $n=3$ and $n=4$.

\begin{table}
{{\bf Table 6}\,\, Values of $|d_{k\, k\, k}|^{1/3k}$ and $|d_{k\,k\, k\, k}|^{1/4k}$
for  values of $m$ in the Taylor expansions}
\renewcommand{\arraystretch}{1}
\begin{tabular}{|c|c|c|c|c|c|c|c|c|}\hline
$k$ & 10 & 50 & 100 & 500 & 750 & 1000 & 1500 & 2000 \\ \hline
$|d_{k\, k\, k}|^{1/3k}$ &2.6596 &2.8977 &2.9416 &2.9850 &2.9895 &2.9918 &2.9943 &2.9956 \\ \hline
$|d_{k\, k\, k\, k}|^{1/4k}$ &3.4819 &3.8443 &3.9112 &3.9773 &3.9840 &3.9876 &3.9913 &3.9933 \\ \hline
\end{tabular}
\end{table}

As can be seen from Table 6 and equation (\ref{buena?}),
$1/R=3$, in dimension 3, and $1/R=4$,  in dimension 4.

\vskip0.1cm
\noindent {\bf 4.2 Solving multivariate differential equations  using Taylor expansions.}
\vskip0.1cm

\noindent {\bf Example 4.2.1}  Let us assume that we want to solve the bivariate differential equation

\begin{equation}\label{bivequ}
\frac{\partial}{\partial x} f(x,y)= \frac{\partial}{\partial y} f(x,y)\quad\mbox{for every}\quad (x,y)\in A,
\end{equation} 
where $A\subset \mathbb{R}^2$ is an open subset.
If we try to find a general solution of this equation, the package Mathematica provides a
general solution of the form $g(x+y)$ where $g$ is a real function, that we will assume to be infinitely
differentiable in an open set $B\subset\mathbb{R}$, such that $C\subset B$, where $C=\{x+y\,|\, (x,y)\in A\}$. 
Let us assume that we have a fixed point  $(x_0,y_0)\in A$ such that $x_0+y_0\in B$, and we will assume that the initial
conditions of the equation (\ref{bivequ}) are $f(x_0,y_0)=b_0$ and 
$\frac{\partial}{\partial x} f(x_0,y_0)= \frac{\partial}{\partial y} f(x_0,y_0)=b_1$.
So, if we know that $f(x,y)=g(x+y)$ for every $(x,y)\in A$, for some $g$ which is unknown, 
then we would have  from (\ref{bivequ}) that $f(x_0,y_0)=g(x_0+y_0)$ and 

\begin{equation}\label{realbivequ}
\frac{\partial^{(1)}}{\partial x^{1}} f(x_0,y_0)= g^{(1)}(x_0+y_0)=b_1=
\frac{\partial^{(1)}}{\partial y^1} f(x_0,y_0).
\end{equation}
Of course we also have that $\frac{\partial^{(1)}}{\partial x^{1}} f(x,y)= g^{(1)}(x+y)$ and
$\frac{\partial^{(1)}}{\partial y^{1}} f(x,y)=g^{(1)}(x+y)$ for every $(x,y)\in A$.

It is obvious from (\ref{realbivequ}) that
$$
\frac{\partial^{(2)}}{\partial x^2}f(x,y)=
\frac{\partial^{(2)}}{\partial y^1\partial x^1} f(x,y)= 
\frac{\partial^{(2)}}{\partial x^1\partial y^1} f(x,y)=
\frac{\partial^{(2)}}{\partial y^2}f(x,y),
$$
for every $(x,y)\in A$, so from the initial conditions and the equation after (\ref{realbivequ}) we have that

$$
\frac{\partial^{(2)}}{\partial x^2}f(x_0,y_0)=
\frac{\partial^{(2)}}{\partial y^1\partial x^1} f(x_0,y_0)= 
\frac{\partial^{(2)}}{\partial x^1\partial y^1} f(x_0,y_0)=
\frac{\partial^{(2)}}{\partial y^2}f(x_0,y_0)= g^{(2)}(x_0+y_0).
$$

Now for every $k>2$ and for every $0\leq s\leq k$ we have that

\begin{equation}\label{derivgral}
\frac{\partial^{(k)}}{\partial x^{(k-s)}\partial y^s}f(x,y)=g^{(k)}(x+y)
=\frac{\partial^{(k)}}{\partial y^{(k-s)} y\partial x^s}f(x,y)
\quad\mbox{for every}\quad (x,y)\in A,
\end{equation}
and from (\ref{derivgral}) it follows that for every $k\geq 1$ and for every $0\leq s\leq k$

\begin{equation}\label{derix0y0}
\frac{\partial^{(k)}}{\partial x^{(k-s)}\partial y^s}f(x_0,y_0)=g^{(k)}(x_0+y_0)
=\frac{\partial^{(k)}}{\partial y^{(k-s)} y\partial x^s}f(x_0,y_0).
\end{equation}

Now, from equation (\ref{taylorbi}) we have that if $f(x,y)$ is infinitely differentiable with respect to
$x$ and $y$, we hav that its Taylor series or McClaurin series around $(x_0,y_0)$ is given by

\begin{equation}\label{MacCbi}
f(x,y)=\sum_{j=0}^{\infty}\sum_{k=0}^{\infty}\frac{\frac{\partial^{(j+k)}}{\partial^{(j)} \partial^{(k)}}f(x_0,y_0)}
{j! k!}(x-x_0)^j (y-y_0)^k.
\end{equation}
In equation (\ref{MacCbi}) we can assume that $f$ is infinitely differentiable in an open set $C\subset \mathbb{R}^2$,
$(x_0,y_0)\in C$ and there exists $r>0$ such that the open ball of radius $r$ and center $(x_0,y_0)$,
denoted by $B_d((x_0,y_0);r)$, where $d$ is the Euclidean distance in $\mathbb{R}^2$, is
included in $C$, and  (\ref{MacCbi}) holds in $B_d((x_0,y_0);r)$.

If we use equations (\ref{derix0y0}) and (\ref{MacCbi}) and we define $z_0=x_0+y_0$,
 we have that for every $(x,y)\in B_d((x_0,y_0);r).$ Let 

\begin{eqnarray}\label{realg}
f(x,y)& =& f(x_0,y_0) \nonumber\\
&+& \left\{ \frac{\partial^{(1)}}{\partial x^1}f(x_0,y_0) \frac{(x-x_0)^{1}}{1!} \frac{(y-y_0)^{0}}{0!}
+\frac{\partial^{(1)}}{\partial y^1}f(x_0,y_0) \frac{(x-x_0)^{0}}{0!} \frac{(y-y_0)^{1}}{1!}\right\} \nonumber\\
&+& \left\{ \frac{\partial^{(2)}}{\partial x^2}f(x_0,y_0)\frac{(x-x_0)^2}{2!}\frac{(y-y_0)^0}{0!}
+\frac{\partial^{(2)}}{\partial x^1\partial y^1}f(x_0,y_0) \frac{(x-x_0)^1}{1!}\frac{(y-y_0)^1}{1!}\right.\nonumber\\
& & +\left. \frac{\partial^{(2)}}{\partial y^2}f(x_0,y_0) \frac{(x-x_0)^0}{0!}\frac{(y-y_0)^2}{2!}\right\} + \cdots \nonumber\\
&= & g^{(0)}(z_0) +\left\{ \left( \frac{(x-x_0)^1}{1!} + \frac{(y-y_0)^1}{1!}\right) g^{(1)}(z_0)\right\} \nonumber\\
& & + \left\{ \left( \frac{(x-x_0)^2}{2!} + \frac{(x-x_0)^1}{1!}\frac{(y-y_0)^1}{1!}
   + \frac{(y-y_0)^2}{2!}\right) g^{(2)}(z_0)\right\}  +\cdots	
\end{eqnarray}
Let $z=x+y$ for every $(x,y)\in C$. Observe that for every $n\geq 0$, 
$$
(z-z_0)^n =((x-x_0)+(y-y_0))^n=\sum_{k=0}^{n}\frac{n!}{k! (n-k)!}(y-y_0)^k (x-x_0)^{(n-k)!}.
$$
Therefore, $ (z-z_0)/n!= \sum_{k=0}^{n} \frac{(y-y_0)^k}{k!} \frac{(x-x_0)^{n-k}}{(n-k)!}$ for every $n\geq 0$.
Hence, using equation (\ref{realg}) and the univariate Taylor series, we get,

$$
f(x,y)=g^{(0)}(z_0)+ g^{(1)}(z_0)\frac{(z-z_0)^1}{1!}+g^{(2)}(z_0)\frac{(z-z_0)^2}{2!}+\cdots=g(z)
$$
for every $(x,y)\in B_d((x_0,y_0);r)$.

Of course in this case the solution of equation (\ref{bivequ}) is not unique since we can select
the univariate function $g$ just by asking that $g$ be infinitely differentiable in a non empty open set
$B\subset \mathbb{R}$. So, in this case the inital conditions do not provide a unique solution.

\vskip .3cm
\noindent {\bf Example 4.2.2} Let us assume that we want to solve the bivariate differential equations

\begin{equation}\label{gh}
\frac{\partial^{(1)}}{\partial x^1}f(x,y)= g^{(0)}(x) f(x,y) \quad \mbox{and}\quad
\frac{\partial^{(1)}}{\partial y^1}f(x,y)= h^{(0)}(y) f(x,y),
\end{equation}
where $f(x_0,y_0)=1$ and  $f:A\rightarrow \mathbb{R}$ is an infinitely differentiable function with 
$A\subset\mathbb{R}^2$  an open set such that
$(x_0,y_0)\in A.$ We will assume that $g^{(0)}$ and $h^{(0)}$ are real functions which are
infinitely differentiable around $x=x_0$ and $y=y_0$.

\noindent {\bf Lemma 5}
{\it The solution to equation (\ref{gh})  can be found using the bivariate Taylor series of the function
$f(x,y),$ and it is given by 
\begin{equation}\label{finalsolution}
f(x,y)= C\cdot e^{\int_{x_0}^x g(t)dt} \cdot e^{\int_{y_0}^{y} h(t)dt},
\end{equation}
where $C$ is a constant.}

\noindent {\bf Proof:}   First, we observe that using equations (\ref{gh})  $f$ satisfies that 
$\frac{\partial^{(1)}}{\partial x^1}f(x_0,y_0)= g^{(0)}(x_0) f(x_0,y_0)=g^{(0)}(x_0)$ and
$\frac{\partial^{(1)}}{\partial y^1}f(x_0,y_0)= h^{(0)}(y_0) f(x_0,y_0)=h^{(0)}(y_0)$.

Second, we will see that $f$ satisfies that for every integer $n\geq 0$
\begin{equation}\label{deriv-enex}
\frac{\partial^{(n+1)}}{\partial x^{n+1}}f(x,y)=\sum_{k=0}^{n}
\left(\begin{array}{c}
n \\
k
\end{array}\right) g^{(n-k)}(x) \frac{\partial^{(k)}}{\partial x^{k}} f(x,y)
\end{equation}
It is clear that for $n=0$, we obtain (\ref{deriv-enex}). Let us assume 
the induction hypothesis, that is,  (\ref{deriv-enex}) holds for $k\leq n$. We have to proof (\ref{deriv-enex})
for $n+1$.

\begin{eqnarray}\label{jijo}
\frac{\partial^{(n+1)}}{\partial x^{n+1}}f(x,y) & = & \frac{\partial^{(1)}}{\partial x^1}
\left\{\frac{\partial^{(n)}}{\partial x^{n}} f(x,y)\right\}  \nonumber \\
& = & \frac{\partial^{(1)}}{\partial x^{1}}
\left\{\sum_{k=0}^{n-1}
\left(\begin{array}{c}
n-1 \\
k
\end{array}\right)  g^{((n-1)-k)}(x) \frac{\partial^{(k)}}{\partial x^{k}} f(x,y)\right\}\nonumber \\
& = & \left(\begin{array}{c}
n-1 \\
0
\end{array}\right)\left[ g^{(n)}(x)f(x,y)+g^{(n-1)}(x)\frac{\partial^{(1)}}{\partial x^1}f(x,y)\right] \nonumber\\
& & + \left(\begin{array}{c}
n-1 \\
1
\end{array}\right)\left[ g^{(n-1)}(x)\frac{\partial^{(1)}}{\partial x^1}f(x,y)+
g^{(n-2)}(x)\frac{\partial^{(2)}}{\partial x^2}f(x,y)\right] \nonumber\\
& & + \quad\quad\quad\vdots \nonumber\\
& & +\left(\begin{array}{c}
n-1 \\
n-1
\end{array}\right)\left[ g^{(1)}(x)\frac{\partial^{(n-1)}}{\partial x^{n-1}}f(x,y)+
g^{(0)}(x)\frac{\partial^{(n)}}{\partial x^n}f(x,y)\right] \nonumber\\
& = & \left(\begin{array}{c}
n \\
0
\end{array}\right) g^{(n)}(x) f(x,y) + \left(\begin{array}{c}
n \\
1
\end{array}\right) g^{(n-1)}(x)\frac{\partial^{(1)}}{\partial x^1} f(x,y) \nonumber\\
& & +\quad\quad\quad \vdots \nonumber \\
& & + \left(\begin{array}{c}
n \\
n-1
\end{array}\right) 
g^{(1)}(x)\frac{\partial^{(n-1)}}{\partial x^{n-1}} f(x,y) +
\left(\begin{array}{c}
n \\
n
\end{array}\right) g^{(0)}(x)\frac{\partial^{(n)}}{\partial x^n} f(x,y) \nonumber\\
& =& \sum_{k=0}^{n} \left(\begin{array}{c}
n \\
k
\end{array}\right) g^{(n-k)}(x)\frac{\partial^{(k)}}{\partial x^k} f(x,y)
\end{eqnarray}
In equation (\ref{jijo}), we used the well known combinatorial formula
$$ \left(\begin{array}{c}
n-1 \\
k
\end{array}\right)+
\left(\begin{array}{c}
n-1 \\
k+1
\end{array}\right)=\left(\begin{array}{c}
n \\
k+1
\end{array}\right)\quad\mbox{and}\quad \left(\begin{array}{c}
n \\
0
\end{array}\right)=1,$$
for every $0\leq k\leq n-1$ and for every $n\geq 1$. Of course, we will have a similar result in 
equation (\ref{jijo}) for $y$ if we interchange
$x$ by $y$ and $g$ by $h$.

From Clairaut's Theorem, since $f$, $g$ and $h$ are infinitely differentiable
with continuous partials we know that all partials 
of the same order are all equal,
that is, for every $n\geq 2$

\begin{eqnarray}\label{1y(n-1)x}
\frac{\partial^{(n)}}{\partial y^1 \partial x^{n-1}} f(x,y)
& =&
\frac{\partial^{(n)}}{\partial x^1 \partial y^1 \partial x^{n-2}} f(x,y)\nonumber\\
& = &
\vdots \nonumber \\ 
& =& \frac{\partial^{(n)}}{\partial x^{n-2}\partial y^1 \partial x^1}f(x,y)\nonumber \\
& =&
\frac{\partial^{(n)}}{\partial x^{n-1}\partial y^1}f(x,y)
\end{eqnarray}
Of course,  (\ref{1y(n-1)x}) can be generalized to $k$ partials with respect to $y$ and to
$n-k$ partials with respect to $x$, in any possible order, for every $1\leq k\leq n-1$. So

\begin{equation}\label{ky(n-k)x}
\frac{\partial^{(n)}}{\partial y^k \partial x^{n-k}} f(x,y)=
\frac{\partial^{(n)}}{\partial y^{k-1}\partial x^1\partial y^1\partial x^{n-k-1}}f(x,y)=\cdots=
\frac{\partial^{(n)}}{\partial x^{n-k} \partial y^{k}} f(x,y).
\end{equation}

Therefore, using (\ref{1y(n-1)x}), we have that
\begin{eqnarray}\label{yunoxn-1}
\frac{\partial^{(n)}}{\partial y^1\partial x^{n-1}}f(x,y) & = & 
\frac{\partial^{(1)}}{\partial y^1}\left( \frac{\partial^{(n-1)}}{\partial x^{n-1}} f(x,y)\right)\nonumber\\
& = & \frac{\partial^{(1)}}{\partial y^1} \left(\sum_{k=0}^{n-2} \left(\begin{array}{c}
n-2\\
k
\end{array}\right)
g^{(n-2-k)}(x) \frac{\partial^{(k)}}{\partial x^k} f(x,y)\right)\nonumber\\
& = & \sum_{k=0}^{n-2} \left(\begin{array}{c}
n-2\\
k
\end{array}\right)
g^{(n-2-k)}(x) \frac{\partial^{(k)}}{\partial x^k} h(y)f(x,y)\nonumber\\
& = & h(y)\cdot \sum_{k=0}^{n-2} \left(\begin{array}{c}
n-2\\
k
\end{array}\right)
g^{(n-2-k)}(x) \frac{\partial^{(k)}}{\partial x^k} f(x,y)\nonumber\\
& = & h(y) \cdot\frac{\partial^{(n-1)}}{\partial x^{n-1}} f(x,y)
\end{eqnarray}

Evaluating (\ref{yunoxn-1}) at $(x,y)=(x_0,y_0)$ we obtain that

$$
\frac{\partial^{(n)}}{\partial y^1\partial x^{n-1}}f(x_0,y_0)=h(y_0)\cdot \frac{\partial^{(n-1)}}{\partial x^{n-1}} f(x_0,y_0)
=\frac{\partial^{(1)}}{\partial y^1} f(x_0,y_0)\cdot\frac{\partial^{(n-1)}}{\partial x^{n-1}} f(x_0,y_0).
$$
We want to prove that for any $n\geq 1$ and for every $0\leq k\leq n$ we have that

\begin{equation}\label{constantsnk}
\frac{\partial^{(n)}}{\partial y^k\partial x^{n-k}}f(x_0,y_0)
=\frac{\partial^{(k)}}{\partial y^k}f(x_0,y_0)\cdot 
\frac{\partial^{(n-k)}}{\partial x^{n-k}}f(x_0,y_0).
\end{equation}
Using (\ref{ky(n-k)x}) we have that (\ref{constantsnk}) holds when we take $k$ partial derivatives for $y$
and $n-k$ partial derivatives for $x$, for every $1\leq k\leq n-1$, in any order.

\noindent Finally, it is not difficult to see that the general solution of equation (\ref{gh}) is given by
the equation (\ref{finalsolution}).  \qed

Here we will study an example where we will compare the Taylor Series obtained from  (\ref{gh})
to the one obtained  by using some code from the package Mathematika.

\noindent {\bf Example 4.2.3} Let us consider the differential equation (\ref{gh}) where
$(x_0,y_0)=(0,\pi/2)$, $g(t) = \sin(t)$ and $h(t)=\cos(t)$ for every $t\in\mathbb{R}$.
Here, we observe that

$$ \int_{x_0}^{x} g(t)dt= \int_{0}^{x}\sin(t)dt=-\cos(t)|_{0}^{x}=1-\cos(x)$$
and

$$ \int_{y_0}^{y} h(t)dt = \int_{\pi/2}^{y} \cos(t) dt=\sin(t)|_{\pi/2}^{y} =\sin(y)-1.$$
Therefore, if $f(x,y)$ is given in equation (\ref{finalsolution}) with $C=1$, we get

$$ f(x,y) =e^{1-\cos(x)}\cdot e^{\sin(y)-1} =e^{\sin(y)-\cos(x)},$$
for every $(x,y)\in\mathbb{R}^2$, hence,

$$ f(x_0,y_0)=f(0,\pi/2)= e^{\sin(\pi/2)-\cos(0)}= e^{1-1}=e^{0}=1.$$

The constants in the Taylor expansion of $f$ are given by  the partial derivatives evaluated
at $(x_0,y_0)=(0,\pi/2),$ so  using (\ref{gh})

$$
\frac{\partial^{(1)}}{\partial x^1}f(x_0,y_0)=g^{(0)} (x_0)f(x_0,y_0)=
\sin(0)f(0,\pi/2)=0\cdot 1=0
$$
and 

$$
\frac{\partial^{(1)}}{\partial y^1}f(x_0,y_0)=h^{(0)} (y_0)f(x_0,y_0)=\cos(\pi/2)f(0,\pi/2)=0\cdot 1=0.
$$
 We also have
 
\begin{eqnarray}
\frac{\partial^{(2)}}{\partial x^2}f(x_0,y_0)& = & g^{(1)} (x_0)f(x_0,y_0)+ g^{(0)}(x_0)
\frac{\partial^{(1)}}{\partial^{x^1}}f(x_0,y_0) \nonumber\\
& = & \cos(0)f(0,\pi/2) + \sin(0) \frac{\partial ^{(1)}}{\partial x^1}f(0,\pi/2) \nonumber\\
& = & 1\cdot 1+0\cdot 0=1 \nonumber
\end{eqnarray}
and
\begin{eqnarray}
\frac{\partial^{(2)}}{\partial y^2}f(x_0,y_0)& = & h^{(1)} (y_0)f(x_0,y_0)+ h^{(0)}(x_0)
\frac{\partial^{(1)}}{\partial^{y^1}}f(x_0,y_0) \nonumber\\
& = & -\sin(\pi/2)f(0,\pi/2) + \cos(\pi/2) \frac{\partial ^{(1)}}{\partial y^1}f(0,\pi/2) \nonumber\\
& = & -1\cdot 1+0\cdot 0=-1, \nonumber
\end{eqnarray}
Besides,

\begin{eqnarray}
\frac{\partial^{(3)}}{\partial x^3}f(x_0,y_0)& = & g^{(2)} (x_0)f(x_0,y_0)+ 2g^{(1)}(x_0)
\frac{\partial^{(1)}}{\partial^{x^1}}f(x_0,y_0) +g^{(0)}(x_0)\frac{\partial^{(2)}}{\partial x^2}f(x_0,y_0)\nonumber\\
& = & -\sin(0)f(0,\pi/2) + 2\cos(0) \frac{\partial ^{(1)}}{\partial x^1}f(0,\pi/2) +
\sin(0) \frac{\partial^{(2)}}{\partial x^2}f(0,\pi/2) \nonumber\\
& = & -0\cdot 1+2\cdot 1\cdot 0 +0\cdot 1=0, \nonumber
\end{eqnarray}
and
\begin{eqnarray}
\frac{\partial^{(3)}}{\partial y^3}f(x_0,y_0)& = & h^{(2)} (y_0)f(x_0,y_0)+ 2h^{(1)}(y_0)
\frac{\partial^{(1)}}{\partial^{y^1}}f(x_0,y_0) +h^{(0)}(y_0)\frac{\partial^{(2)}}{\partial y^2}f(x_0,y_0)\nonumber\\
& = & -\cos(\pi/2)f(0,\pi/2) - 2\sin(\pi/2) \frac{\partial ^{(1)}}{\partial y^1}f(0,\pi/2) \nonumber\\
&+& \cos(\pi/2) \frac{\partial^{(2)}}{\partial y^2}f(0,\pi/2)  =  -0\cdot 1-2\cdot(-1)\cdot 0 +0\cdot(-1)=0, \nonumber
\end{eqnarray}

Also,
\begin{eqnarray}
\frac{\partial^{(4)}}{\partial x^4}f(x_0,y_0)& = & g^{(3)} (x_0)f(x_0,y_0)+ 3g^{(2)}(x_0)
\frac{\partial^{(1)}}{\partial^{x^1}}f(x_0,y_0) +3g^{(1)}(x_0)\frac{\partial^{(2)}}{\partial x^2}f(x_0,y_0)\nonumber\\
& &+ g^{(0)}(x_0)\frac{\partial^{(3)}}{\partial x^3} f(x_0, y_0)\nonumber\\
& = & -1\cdot 1 + 3\cdot 0\cdot \frac{\partial ^{(1)}}{\partial x^1}f(0,\pi/2) +
3\cdot 1\cdot \frac{\partial^{(2)}}{\partial x^2}f(0,\pi/2) \nonumber\\
&+& 0\cdot \frac{\partial^{(3)}}{\partial x^3}f(0,\pi/2) 
 =  -1+3\cdot 1\cdot 1=2, \nonumber
\end{eqnarray}
and

\begin{eqnarray}
\frac{\partial^{(4)}}{\partial y^4}f(x_0,y_0)& = & h^{(3)} (x_0)f(x_0,y_0)+ 3h^{(2)}(x_0)
\frac{\partial^{(1)}}{\partial {y^1}}f(x_0,y_0) +3h^{(1)}(x_0)\frac{\partial^{(2)}}{\partial y^2}f(x_0,y_0)\nonumber\\
& &+ h^{(0)}(x_0)\frac{\partial^{(3)}}{\partial y^3} f(x_0, y_0)\nonumber\\
& = & 1\cdot 1 + 3\cdot 0\cdot \frac{\partial ^{(1)}}{\partial y^1}f(0,\pi/2) +
3\cdot (-1)\cdot \frac{\partial^{(2)}}{\partial y^2}f(0,\pi/2) \nonumber\\
&+& 0\cdot \frac{\partial^{(3)}}{\partial y^3}f(0,\pi/2)
 =  1+3\cdot 1\cdot 1=4, \nonumber
\end{eqnarray}

\noindent Hence, the approximation of $f(x,y)$ around $(x_0,y_0)=(0,\pi/2)$ using a Taylor polynomial with 
$n=m=4$ with 16 terms with $C_{1,j}=\frac{\partial ^{(j)}}{\partial x^j} f(0,\pi/2)$ and
$C_{2,k}=\frac{\partial^{(k)}}{\partial y^k}f(0,\pi/2)$, we get that

\begin{eqnarray}
f(0,\pi/2)&\approx&\sum_{j=0}^{4} \sum_{k=0}^{4} C_{1,j}\cdot  C_{2,k} \cdot \frac{(x-0)^j}{j!}\cdot \frac{(y-\pi/2)^k}{k!}\nonumber\\
&=& \left( \sum_{j=1}^{4}C_{1,j}\frac{x^j}{j!}\right)\cdot \left( \sum_{k=0}^{4} C_{2,k} \frac {(y-\pi/2)^k}{k!}
\right)\nonumber\\
&=& \left[ C_{1,0}+C_{1,1}x+ C_{1,2}\frac{x^2}{2!} + C_{1,3}\frac{x^3}{3!}+ C_{1,4}\frac{x^4}{4!}\right] \nonumber\\
& & \cdot\left[ C_{2,0}+C_{2,1}(y-\pi/2) + \dots
    +C_{2,4}\frac{(y-\pi/2)^4}{4!}\right]\nonumber\\		
&=& \left[ 1+0\cdot x+ 1\cdot \frac{x^2}{2}+0\cdot \frac{x^3}{6}+ 2\cdot\frac {x^4}{24}	\right]\nonumber\\
& & \cdot \left[ 1+0\cdot (y-\pi/2) -\frac{(y-\pi/2)^2}{2}+0\cdot \frac{(y-\pi/2)^3}{6}+4\cdot\frac{(y-\pi/2)^4}{24}
\right]\nonumber\\
& = & \left[1+\frac{x^2}{2}+\frac{x^4}{12}\right]\cdot \left[ 1-\frac{(y-\pi/2)^2}{2}+\frac{(y-\pi/2)^4}{6}\right]=
P_4(x,y). \nonumber
\end{eqnarray}
If we evaluate $P_4(1,1)$ we obtain

\begin{eqnarray}
P_4(1,1)&=& \left[ 1+\frac{1}{2}+\frac{1}{12}\right] \cdot \left[1-\frac{(1-\pi/2)^2}{2}+\frac{(1-\pi/2)^4}{6}\right]\nonumber\\
& = & \frac{19}{12}\cdot \left[ 1-\frac{1}{2} (1-\pi/2)^2+\frac{1}{6}(1-\pi/2)^4 \right]\nonumber\\
& =& \frac{19}{12}-\frac{19}{24} (1-\pi/2)^2 +\frac{19}{72} (1-\pi/2)^4. \nonumber
\end{eqnarray}
This is the result obtained with the aid of  Mathematica by  specifying  $(x_0,y_0)=(0,\pi/2),$ $g(t) = \sin(t)$ and $h(t)=\cos(t),$
which corresponds  to  approximate  the solution of the equation (\ref{gh}).

\setcounter{equation}{58}
\vskip0.1cm
\noindent {\bf Remark 4.2.4} After a long time of searching for results about the Cauchy-Hadamard theorems for the {\bf radius of convergence} $R,$
of a Taylor's series in the multivariate setting,  we found few results in the literature.  The only reasonable
result about this case that we found, is a Note by  Ulrik Skre Fjordholm (see \cite{FJ}) whose title is {\it A note on multiindices},
from 2020.  This note  approaches smooth functions $f:\mathbb{R}^d\rightarrow \mathbb{R},$ in the $d$-multivariate case of Taylor's 
expansion by using multiindices $\underline{\alpha}\in \mathbb{N}_0^d$, where $\mathbb{N}_0$ is the set of nonnegative integers.
If we let $\underline{\alpha}=(\alpha_1,\alpha_2,\ldots ,\alpha_d)$, where $\alpha_i\in\mathbb{N}_0$ for every 
$i\in \{1,2, \ldots ,d\}$ and defining $|\underline{\alpha}|=\sum_{i=1}^{d} \alpha_i$ and we define
\begin{equation}
\left(\begin{array}{c}
|\underline{\alpha}| \nonumber\\
\underline{\alpha} \nonumber
\end{array}\right) =\frac{|\alpha|!}{\alpha!}=\frac{(\alpha_1+\cdots +\alpha_d)!}{\alpha_1!\cdots \alpha_d!}.
\end{equation}
Given two multiindices $\underline{\alpha}$ and $\underline{\beta}$, 
$\underline{\alpha}\leq \underline{\beta}$ if and only if
$\alpha_i\leq \beta_i$ for every $i\in \{1,2,\ldots ,d\}$. If $\underline{x}=(x_,\ldots ,x_d)\in\mathbb{R}^d$,
 then $\underline{x}^{\underline{\alpha}}= x_1^{\alpha_1}\cdots x_d^{\alpha_d}$ and the derivatives of $f$
are given by
$$
f^{(\underline{\alpha})}(\underline{x})=\frac{\partial^{|\underline{\alpha}|}f}
{\partial x_1^{\alpha_1}\cdots \partial x_d^{\alpha_d}} (\underline{x}).
$$
Then he gives a proof of the Taylor's formula around a point $\underline{z}\in\mathbb{R}^d$
by defining
$g(t)=f(\underline{z}+t(\underline{x}-\underline{z}))$ for every $t \in \mathbb{R}$.
 Finally he proves that the radius of convergence can be written as
$$
R = \frac{1}{\limsup_{|\underline{\alpha}|\rightarrow \infty} |C_{\underline{\alpha}}|^{1/|\underline{\alpha}|}},
$$ 
where $C_{\underline{\alpha}}$ is given in (\ref{buena?})
He also provides the proof of the Cauchy-Hadamard's Theorem.

\vskip0.1cm
\noindent {\bf 5 Conclusion.}
\vskip0.1cm

In this work, we illustrate the use of Taylor series to find solutions of differential equations corresponding to a number of classic examples, also  some multivariate cases within
the framework of partial differential equations are studied.   In applications it is quite common to use a finite Taylor Polynomial as a reasonable approximation to the solution, however
we propose to consider the entire series as a solution which in some cases allows us to find an analytic (closed-form) expression for it.  When this analytic form is not available,
the Cauchy-Hadamard theorems can provide a convergence radius, which in turn determines a domain where the series is defined  as a solution to the differential equation.

\end{document}